# UDL-RT: A unified differentiable all-sky microwave radiative transfer model with learnable parameterizations

Zeting Li[1,2,3], Wei Han[2,3], Jie Peng[1], Xuewei Zhang[4], Xiaofeng Wang[1], Xueshun Shen[3]

[1]Shanghai Meteorological Equipment Support Center, China Meteorological Administration, Shanghai 200050, China.

[2]State Key Laboratory of Severe Weather, Chinese Academy of Meteorological Sciences, Beijing 100081, China.

[3]CMA Earth System Modeling and Prediction Centre (CEMC), Beijing 100081, China.

[4]Key Laboratory of Meteorological Disaster, Ministry of Education (KLME)/Joint International Research Laboratory of Climate and Environment Change (ILCEC)/Collaborative Innovation Center on Forecast and Evaluation of Meteorological Disasters (CIC-FEMD), Nanjing University of Information Science & Technology, Nanjing 210044, China.

*Correspondence to*: Wei Han (hanwei@cma.gov.cn)

**Abstract.** Radiance-based retrievals and data assimilation require radiative transfer models (RTMs) to simulate radiances and quantify their sensitivities to atmospheric and surface conditions. Conventional RTMs often rely on separately maintained forward, tangent-linear, and adjoint codes, complicating model development and extension. Here, we present the Unified Differentiable and Learnable Radiative Transfer model (UDL-RT), a PyTorch-based framework for all-sky microwave radiative transfer that couples neural-network (NN) parameterizations of gaseous absorption, hydrometeor bulk scattering properties (BSPs), and ocean surface emissivity with an explicit radiative transfer solver in a unified differentiable computational graph. Automatic differentiation directly computes Jacobians with respect to input variables, reducing the model maintenance burden. The NN parameterizations offer flexible representations of complex nonlinear relationships. For example, the NN gas-absorption parameterization reduces the standard deviation of brightness-temperature (TB) errors by approximately 60-85% in the water-vapour channels compared with linear regression. Moreover, UDL-RT propagates loss gradients to trainable parameters to optimize physical parameterizations against observations, with GPU-accelerated batched computation supporting efficient iterative learning. Using Microwave Humidity Sounder-2 (MWHS-2) observations, we optimize the snow BSP parameterization through gradient-based learning. The optimized parameterization improves agreement between simulated and observed TB distributions and reduces the systematic bias and skewness of observation-minus-background (O-B) departures. Overall, UDL-RT integrates radiance simulation, sensitivity analysis, and parameterization optimization within a unified framework, supporting future remote sensing retrieval and data assimilation applications.

***Keywords***: All-sky radiative transfer; Microwave remote sensing; Automatic differentiation; Neural network; Parameter optimization

**Highlights**:

1. UDL-RT computes all-sky microwave radiances and obtains Jacobians automatically.

2. A hybrid framework couples a physical solver with NN parameterizations.

3. Learnable physical parameterizations can be optimized against observations.

# 1 Introduction

Passive microwave observations are widely used to retrieve atmospheric temperature and humidity, clouds and precipitation, and surface properties, and to constrain atmospheric analyses in numerical weather prediction (NWP) (Boukabara et al., 2011; Li and Han, 2025; Yang et al., 2023). Radiative transfer models (RTMs) simulate microwave radiances from atmospheric and surface conditions, providing a quantitative link between geophysical variables and observations. In retrieval and data assimilation, RTMs are often required not only to simulate radiances but also to quantify how the simulated radiances respond to changes in the state variables. These sensitivities, expressed as Jacobians, are essential for gradient-based methods such as optimal estimation and variational data assimilation (Errico, 1997; Rodgers, 2000).

A variety of microwave RTMs have therefore been developed to support these applications, ranging from flexible research models such as the Atmospheric Radiative Transfer Simulator (ARTS) (Buehler et al., 2018) and Passive and Active Microwave TRAnsfer (PAMTRA) (Mech et al., 2020) to fast operational models such as Radiative Transfer for TOVS (RTTOV) (Saunders et al., 2018), the Community Radiative Transfer Model (CRTM) (Johnson et al., 2023; Liu and Boukabara, 2014), and the Advanced Radiative Transfer Modeling System (ARMS) (Weng et al., 2020). Several RTMs also provide tangent-linear and adjoint models for computing derivatives with respect to state variables (Bauer et al., 2006). However, these derivative models are closely tied to the corresponding forward model. Changes to the forward calculation or extensions to additional variables and model parameters require corresponding updates to the tangent-linear and adjoint models, increasing the development and maintenance burden (Giering and Kaminski, 1998; Liu and Cao, 2019). Meanwhile, the increasing complexity of all-sky simulation and the growing volume of satellite observations place greater computational demands on both forward and derivative calculations (Barlakas et al., 2022; Duruisseau et al., 2017; Eriksson et al., 2020). Together, these considerations motivate computational approaches that can provide derivatives flexibly while maintaining high computational efficiency.

Differentiable computing frameworks such as PyTorch (Paszke et al., 2019) and JAX (Bradbury et al., 2018) offer a potential route to address these requirements by combining automatic differentiation with efficient numerical computation on modern hardware. One application of these frameworks to radiative transfer is the development of neural network (NN) emulators. Rather than explicitly solving the radiative transfer problem, these models learn an end-to-end mapping from atmospheric and surface variables to radiances. Following this approach, NN emulators have been developed to simulate clear-sky microwave radiances (Liang et al., 2022), all-sky microwave radiances (Li et al., 2024a), and hyperspectral infrared radiances (Li et al., 2025). Such emulators can substantially reduce the cost of forward calculations and compute Jacobians via automatic differentiation. However, NNs are generally trained to reproduce RTM outputs rather than their derivatives. Accurate radiance simulations therefore do not necessarily imply reliable Jacobians, which may exhibit local oscillations and deviate from physical-model Jacobians (Aires et al., 1999; Li et al., 2024a).

An alternative is to implement the physical RTM itself within a differentiable computing framework. Because the derivatives are obtained from the same computational graph as the forward calculation, changes to differentiable model components are automatically reflected in the derivatives, avoiding the need to maintain corresponding tangent-linear and adjoint models (Baydin et al., 2018). The framework also allows selected uncertain model components to be optimized using gradient-based methods (Gelbrecht et al., 2023; Li et al., 2024b; Shen et al., 2023; Tsai et al., 2021). For example, when a physical parameterization is represented by an NN, loss gradients can be propagated through the radiative transfer calculation to update its trainable parameters. NeuralGCM adopts a related strategy by coupling a differentiable dynamical core with learned parameterizations (Kochkov et al., 2024). More generally, this approach retains established physical formulations while allowing selected uncertain parameterizations to be learned from data.

Recent studies of differentiable radiative transfer in remote sensing have focused on land-surface applications, including vegetation, soil, and urban scenes (Cao et al., 2025; Qi et al., 2026; Sun et al., 2025; Wang et al., 2024), whereas applications to atmospheric radiative transfer remain relatively limited, particularly for all-sky microwave radiance simulations. Extending this approach to all-sky radiative transfer is challenging because gaseous absorption, hydrometeor optical properties, surface boundary conditions, and the multiple-scattering radiative transfer solver must all be implemented differentiably and coupled consistently within the same forward model. In addition, some physical parameterizations in RTMs remain subject to substantial uncertainty, particularly parameterizations of hydrometeor bulk scattering properties (BSPs) (Geer et al., 2021; Wu et al., 2024). A differentiable RTM enables the parameters of these schemes to be optimized directly against radiance observations using gradient-based methods, rather than relying on computationally expensive searches over discrete parameter combinations.

Here, we present UDL-RT, a PyTorch-based radiative transfer framework that couples an explicit physical solver with learnable neural-network parameterizations within a unified differentiable computational graph. The current implementation supports the application of all-sky simulations for both satellite and ground-based passive microwave observations. Automatic differentiation computes Jacobians with respect to atmospheric and surface variables from the same graph used for forward simulation. This graph also propagates observation-based loss gradients to trainable parameters, allowing selected parameterizations to be optimized within the physical radiative transfer model. UDL-RT supports batched computation on CPUs and GPUs, with GPU parallelism accelerating repeated forward and gradient evaluations for retrieval, data assimilation, and parameterization optimization.

The remainder of this paper is organized as follows. Section 2 presents the formulation and implementation of the UDL-RT model, including the gaseous absorption and hydrometeor BSP parameterizations, surface boundary treatment, radiative transfer solver, and automatic differentiation framework. Section 3 evaluates the accuracy of the gaseous absorption and hydrometeor BSP parameterizations, the all-sky forward simulations and Jacobians, and the computational performance of UDL-RT. Section 4 demonstrates the observation-constrained optimization capability of UDL-RT, using the snow BSP parameterization and Microwave Humidity Sounder-2 (MWHS-2) observations as an example. Section 5 discusses the main findings, current limitations, and directions for future development.

## 2 Model formulation

### 2.1 Model overview

UDL-RT is a one-dimensional differentiable RTM for passive microwave observations. Given atmospheric and surface states together with the observation geometry, UDL-RT is able to simulate channel radiances for both spaceborne and ground-based microwave observations. Atmospheric columns are treated independently under the plane-parallel approximation, with scattering represented using an azimuthally averaged scalar radiative transfer formulation. Thermal emission follows local thermodynamic equilibrium, while solar radiation is neglected.

Under these assumptions, the scalar radiative transfer equation is solved separately for each atmospheric column and sensor channel:

$$\mu \frac{\partial I(\tau,\mu)}{\partial \tau} = I(\tau,\mu) - J(\tau,\mu) \tag{1}$$

where $\tau$ is the vertical extinction optical depth increasing downward from the top of the atmosphere, and $\mu$ is the cosine of the propagation zenith angle. Here, $I$ is the radiance and $J$ is the source function. The source function includes thermal emission and scattering:

$$J(\tau,\mu) = (1-\omega_0)B(T) + \frac{\omega_0}{2}\int_{-1}^{1} P\left(\mu,\mu'\right) I\left(\tau,\mu'\right) \mathrm{d}\mu' \tag{2}$$

where $B(T)$ is the Planck radiance at temperature $T$, $\omega_0$ is the single-scattering albedo (SSA), and $P\left(\mu,\mu'\right)$ is the azimuthally averaged scattering phase function. The SSA gives the fraction of total extinction due to scattering, while $P\left(\mu,\mu'\right)$ describes the redistribution of radiation from direction $\mu'$ into direction $\mu$.

As summarized in Fig. 1, UDL-RT includes parameterizations for gaseous absorption, hydrometeor optical properties, and surface emissivity and reflectance. The NN parameterizations for gaseous absorption and hydrometeor BSPs are hereafter referred to as GasAbs-NN and BSP-NN, respectively. Over ocean surfaces, emissivity is calculated using the SURface Fast Emissivity Model for Ocean (SURFEM-Ocean; Kilic et al., 2023), whereas the Tool to Estimate Land Surface Emissivity from Microwave to Submillimeter Waves (TELSEM2; Wang et al., 2017) is used over land and sea ice. The resulting gaseous optical depths, hydrometeor BSPs, and surface emissivity and reflectance, together with the instrument settings, are supplied to the radiative transfer solver. The solver computes upwelling or downwelling radiances, which are then converted to brightness temperatures (TBs). GasAbs-NN, BSP-NN, SURFEM-Ocean, and the radiative transfer solver form a differentiable computational graph, allowing derivatives of the simulated TB with respect to model inputs and trainable parameters to be obtained through automatic differentiation.

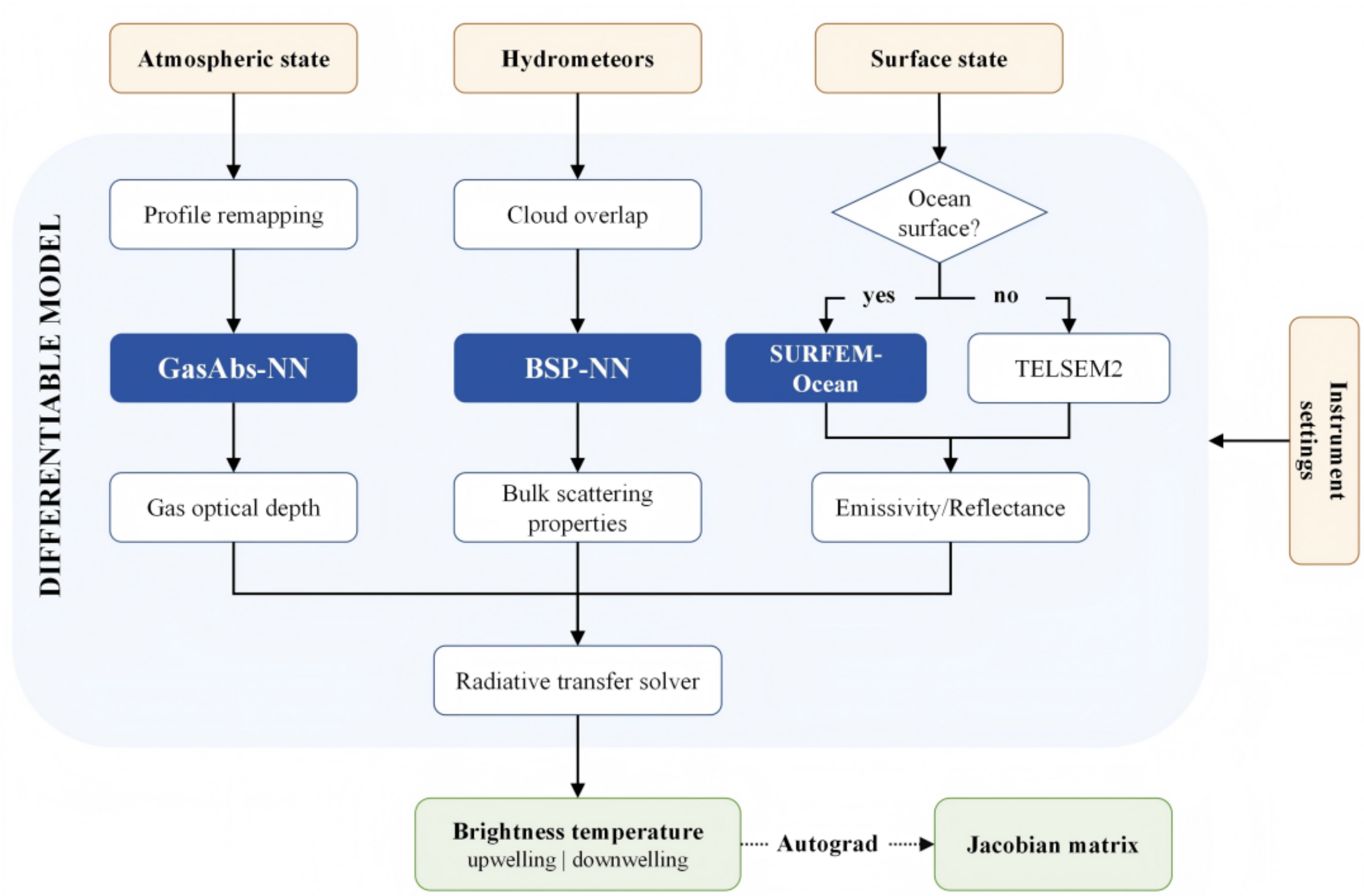

**Fig. 1.** Overview of the UDL-RT framework. Differentiable parameterizations are coupled with an explicit physical radiative transfer solver to simulate brightness temperatures, with the corresponding Jacobians obtained through automatic differentiation.

### 2.2 Gaseous absorption

Gaseous absorption in UDL-RT is represented by channel-dependent layer optical depths. Reference optical depths are obtained from line-by-line (LBL) calculations and used to develop two fast gaseous absorption parameterizations: a predictor-based linear-regression parameterization and an NN parameterization.

Let $\Gamma_{j-1/2}(\nu)$ and $\Gamma_{j+1/2}(\nu)$ denote the monochromatic slant-path transmittances from the top of the atmosphere to the upper and lower boundaries of layer $j$, respectively. For a sensor channel with finite spectral bandwidth, the channel-averaged transmittance at each layer boundary is obtained by integrating the monochromatic transmittance over the normalized spectral response function (SRF):

$$\Gamma_{i,j\pm1/2} = \sum_{m=1}^{M_i} a_{i,m}\,\Gamma_{j\pm1/2}(\nu_{i,m}), \qquad \sum_{m=1}^{M_i} a_{i,m} = 1 \tag{3}$$

where $M_i$ is the number of spectral samples, and $\nu_{i,m}$ and $a_{i,m}$ are the corresponding sampling frequencies and normalized weights for channel $i$. The corresponding slant-path layer optical depth for channel $i$ is

$$\Delta\tau_{i,j}^{\mathrm{gas}} = -\ln\left[\frac{\Gamma_{i,j+1/2}}{\Gamma_{i,j-1/2}}\right] \tag{4}$$

Reference slant optical depths are generated for each profile at six observing angles using the Monochromatic Radiative Transfer Model (MonoRTM) v5.6 (Clough et al., 2005), which performs monochromatic radiative transfer calculations in the microwave and millimeter-wave spectral regions. These optical depths are used as prediction targets for the linear-regression and NN parameterizations described in Sections 2.2.1 and 2.2.2, respectively. Atmospheric profiles are drawn from the 2025 NWP SAF diverse-profile dataset derived from ECMWF Copernicus Atmosphere Monitoring Service (CAMS) 137-level short-range forecasts (Turner, 2025), which contains 100,000 profiles. From this dataset, 4,000 profiles are selected to span a broad range of temperature and humidity conditions, with 3,200 samples used for training and 800 samples for validation. The 83-profile ECMWF diverse dataset (Matricardi, 2008) is also included in the training set, giving a total of 3,283 training profiles. An independent test dataset, described in Section 3.1, is used for the final evaluation.

Both gaseous absorption parameterizations operate on a fixed pressure grid, whereas input atmospheric profiles may be defined on arbitrary vertical grids. The input profiles are therefore interpolated to the fixed grid for the optical-depth calculation, and the resulting optical depths are remapped to the input grid while preserving the total column optical depth. The slant optical depths are then converted to vertical optical depths for use by the radiative transfer solver.

### *2.2.1 Linear-regression parameterization*

UDL-RT implements a linear-regression parameterization based on the mixed-gas and water-vapour predictors described by Saunders et al. (1999). Gaseous absorption is separated into mixed-gas and water-vapour contributions, which are combined to obtain the total gaseous optical depth:

$$\Delta\tau_{i,j}^{\mathrm{gas}} = \Delta\tau_{i,j}^{\mathrm{mix}} + \Delta\tau_{i,j}^{\mathrm{wv}} \tag{5}$$

The layer optical depth of each component is represented as a linear combination of its predictors:

$$\Delta\tau_{i,j}^{\mathrm{mix}} = \sum_{k=1}^{N_{\mathrm{mix}}} C_{i,j,k}^{\mathrm{mix}} X_{j,k}^{\mathrm{mix}}, \qquad \Delta\tau_{i,j}^{\mathrm{wv}} = \sum_{k=1}^{N_{\mathrm{wv}}} C_{i,j,k}^{\mathrm{wv}} X_{j,k}^{\mathrm{wv}} \tag{6}$$

Here, $X_{j,k}$, $C_{i,j,k}$, and $N$ denote the predictor, regression coefficient, and number of predictors, respectively. The regression coefficients are fitted separately for each sensor channel and pressure layer using the MonoRTM reference optical depths described above.

### *2.2.2 Neural-network parameterization*

The linear-regression parameterization is computationally efficient, but nonlinear relationships must be represented through a predefined set of transformed, cross-term, and vertically integrated predictors. NNs can represent more complex nonlinear relationships and directly learn the mapping from atmospheric profile variables to layer optical depths, reducing the dependence on manually designed predictors (Huang et al., 2026; Stegmann et al., 2022).

Because gaseous optical depth is predicted separately for each atmospheric layer, UDL-RT represents the atmospheric profile as an ordered vertical sequence and uses two one-dimensional residual convolutional NNs for mixed-gas and water-vapour absorption. A conventional multilayer perceptron (MLP) applied to the full atmospheric profile typically requires the vertical dimension to be flattened, so the layer structure is not explicitly represented in the network architecture. One-dimensional convolutions instead operate directly along the vertical dimension and preserve the correspondence between atmospheric layers and the predicted layer optical depths. Local convolutions capture information from neighbouring layers, while progressively dilated convolutions expand the receptive field to incorporate broader vertical context within the same atmospheric profile.

The network inputs include pressure, temperature, observing geometry, vertical position, and propagation-path information. The water-vapour network additionally includes humidity and column water-vapour information. Both networks predict log-transformed layer optical depths, reducing their large dynamic range across atmospheric layers and channels.

As illustrated in Fig. 2, each network consists of an input projection, four vertical residual convolutional blocks, and an output projection. Each residual block contains two one-dimensional convolutional layers with a kernel width of five. Dilation factors of 1, 2, 4, and 8 are applied successively across the four blocks to expand the receptive field along the vertical dimension. Both networks use a hidden feature dimension of 80 and contain approximately 260,000 trainable parameters each.

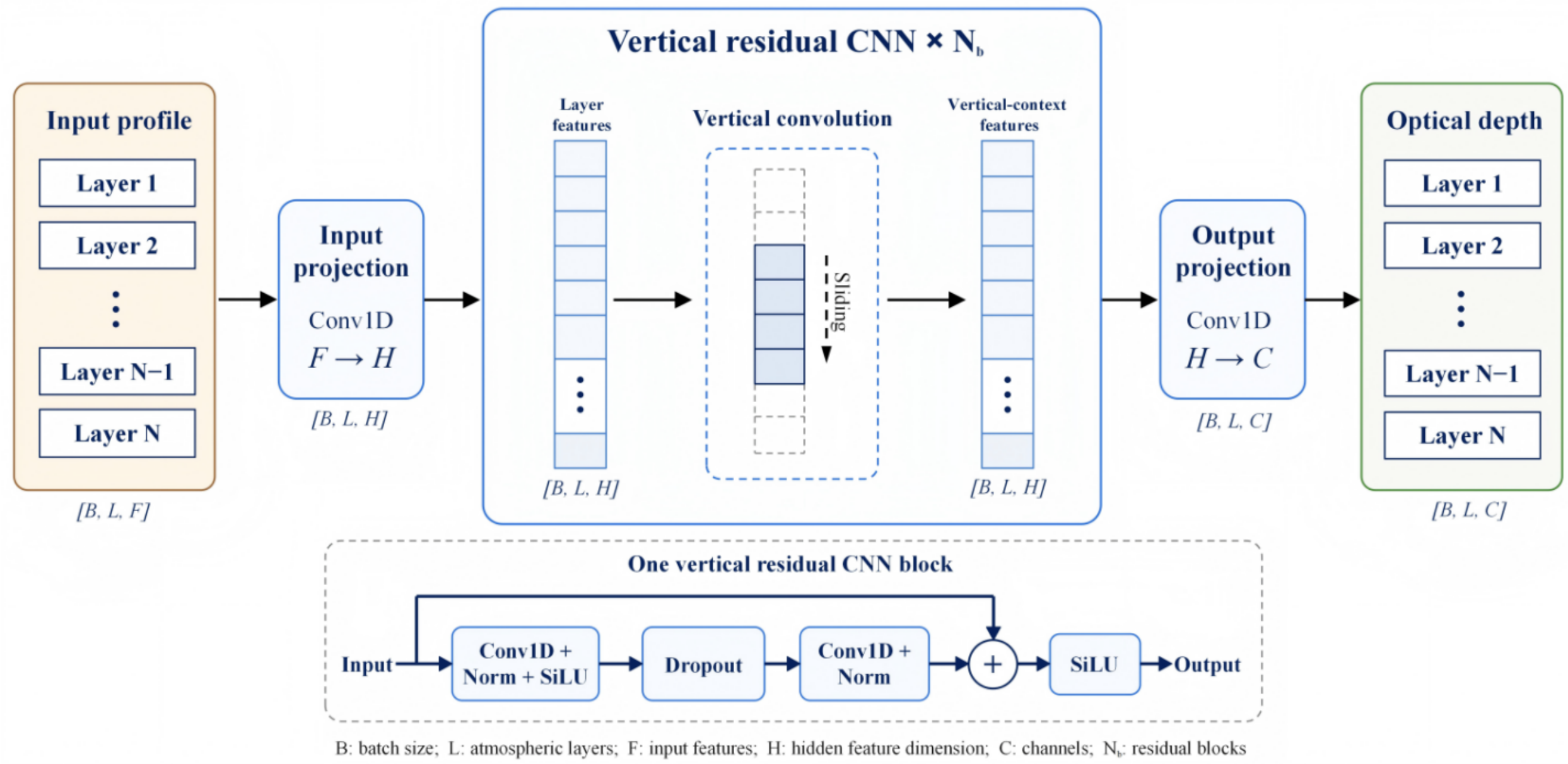

**Fig. 2.** Schematic architecture of the vertical residual convolutional neural network used for gaseous absorption parameterization. The same architecture is used for the mixed-gas and water-vapour networks, with the lower panel illustrating a single residual block.

### 2.3 Hydrometeor optical properties

Under cloudy and precipitating conditions, liquid and frozen hydrometeors absorb and scatter microwave radiation. Electromagnetic scattering calculations provide the optical properties of individual particles, whereas the radiative transfer solver requires BSPs for the hydrometeors in each atmospheric layer. Reference BSPs are calculated by integrating single-particle optical properties over prescribed particle size distributions (PSDs). BSP-NN emulates these integrals at runtime.

Let $n(D)$ denote the PSD, where $D$ is the maximum particle dimension. For a prescribed hydrometeor mass concentration $W$, the PSD is constructed such that its integrated particle mass is consistent with $W$:

$$W = \int_{D_{\min}}^{D_{\max}} m\,(D) n(D)\, \mathrm{d}D \tag{7}$$

where the particle mass is related to size by $m(D) = aD^b$, with $a$ and $b$ determined by the particle model. The resulting PSD, together with the corresponding single-particle optical properties, determines the layer BSPs (Geer et al., 2021).

The bulk extinction coefficient is obtained by integrating the single-particle extinction cross section over the PSD:

$$\kappa_{\text{ext}} = \int_{D_{\min}}^{D_{\max}} C_{\text{ext}}(D) n(D)\, \mathrm{d}D \tag{8}$$

where $C_{\text{ext}}$ is the extinction cross section of an individual particle. Similarly, the scattering coefficient is

$$\kappa_{\text{sca}} = \int_{D_{\min}}^{D_{\max}} C_{\text{sca}}(D) n(D)\, \mathrm{d}D \tag{9}$$

where $C_{\text{sca}}$ is the scattering cross section. The single-scattering albedo is obtained as the ratio of the scattering coefficient to the extinction coefficient. The phase function moments are weighted by the contribution of each particle size to scattering:

$$\beta_{\ell} = \frac{\int_{D_{\min}}^{D_{\max}} C_{\text{sca}}(D) \beta_{\ell}^{\text{single}}(D) n(D)\, \mathrm{d}D}{\kappa_{\text{sca}}}, \qquad \ell = 1, \ldots, N_{\beta} \tag{10}$$

where $\beta_{\ell}^{\text{single}}(D)$ is the $\ell$th Legendre moment of the single-particle phase function. The zeroth moment is normalized to unity, $\beta_0 = 1$. The first moment, $\beta_1 = g$, is the asymmetry parameter and describes the overall preference for forward or backward scattering. Higher-order moments progressively represent finer angular variations of the phase function, including sharper scattering features that cannot be described by the asymmetry parameter alone. UDL-RT retains phase-function moments up to order 15, consistent with the Legendre expansion used for the highest-order 16-stream configuration supported by the radiative transfer solver (Stamnes et al., 1988).

Five hydrometeor species are considered: cloud liquid water, cloud ice, rain, snow, and graupel. The optical properties of liquid particles are calculated using Mie theory. For frozen hydrometeors, the properties of individual particles are taken from the ARTS scattering database (Eriksson et al., 2018), which provides precomputed electromagnetic scattering properties for nonspherical ice particles over a broad frequency range. The particle models and PSDs used for frozen hydrometeors mainly follow the configurations selected by Geer (2021), in which particle shape and PSD assumptions were constrained using near-global microwave observations. The particle models and PSDs adopted for the five hydrometeor species in UDL-RT are summarized in Table 1.

**Table 1.** Microphysical configurations for all hydrometeor types

| Type | Particle shape | $D_{min}$ [m] | $D_{max}$ [m] | $a$ | $b$ | Particle size distribution |
|---|---|---|---|---|---|---|
| Cloud liquid water | Sphere | $1.24\times10^{-6}$ | $5.00\times10^{-2}$ | 523.6 | 3.00 | Petty and Huang (2011), with $\mu$=2.0, $\Lambda$=$2.13\times10^{5}$, $\gamma$=1.0 |
| Rain | Sphere | $1.24\times10^{-6}$ | $5.00\times10^{-2}$ | 523.6 | 3.00 | Marshall and Palmer (1948) |
| Cloud ice water | Large column aggregate | $2.42\times10^{-5}$ | $2.00\times10^{-2}$ | 0.28 | 2.44 | Petty and Huang (2011), with $\mu$=0.0, $\Lambda$=$1.00\times10^{4}$, $\gamma$=1.0 |
| Snow | Large plate aggregate | $1.00\times10^{-4}$ | $2.29\times10^{-2}$ | 0.21 | 2.26 | Field et al. (2007), tropical |
| Graupel | Column type 1 | $1.00\times10^{-4}$ | $1.00\times10^{-2}$ | 0.04 | 2.05 | Field et al. (2007), tropical |

In many fast microwave RTMs, BSPs are precomputed offline for prescribed particle models and PSDs over discrete grids of frequency, temperature, and hydrometeor content. The resulting values are stored in lookup tables and interpolated during the forward calculation. This avoids repeated particle scattering calculations and PSD integrations at runtime, but requires the BSP inputs to be sampled on discrete grids, which can lead to substantial storage requirements as the number and resolution of the input dimensions increase.

Previous studies have shown that NNs can accurately emulate both single-particle and bulk scattering properties while substantially reducing the storage required by conventional scattering databases (Li et al., 2026; Yu et al., 2022). Following this approach, UDL-RT uses a separate residual MLP for each hydrometeor species to parameterize its BSPs over the training frequency range of 1.0-886.4 GHz.

For each hydrometeor species, the network takes temperature, hydrometeor mass concentration, and frequency as inputs. Temperature is linearly scaled, hydrometeor mass concentration is logarithmically transformed and scaled, and frequency is logarithmically transformed and encoded using Fourier features. The transformed variables are then processed by an input projection, $N_b$ residual blocks, and an output projection, as illustrated in Fig. 3. The network predicts the extinction coefficient, single-scattering albedo, and the first 15 phase-function moments. Output transformations ensure a positive extinction coefficient, constrain the single-scattering albedo to values between zero and one, and bound the phase-function moments within [-1, 1].

For satellite radiance simulations using atmospheric fields that represent averages over a model grid cell, cloud fraction is accounted for before the BSPs are calculated. UDL-RT follows Geer et al. (2009), in which an effective cloud fraction is derived from the vertical distributions of cloud fraction and hydrometeor content. The hydrometeor contents in the cloudy column are adjusted accordingly, and clear and cloudy columns are then simulated separately. The resulting radiances are combined using the effective cloud fraction. For ground-based microwave radiometers, this treatment is unnecessary when the supplied profile represents the atmospheric column sampled along the observation path.

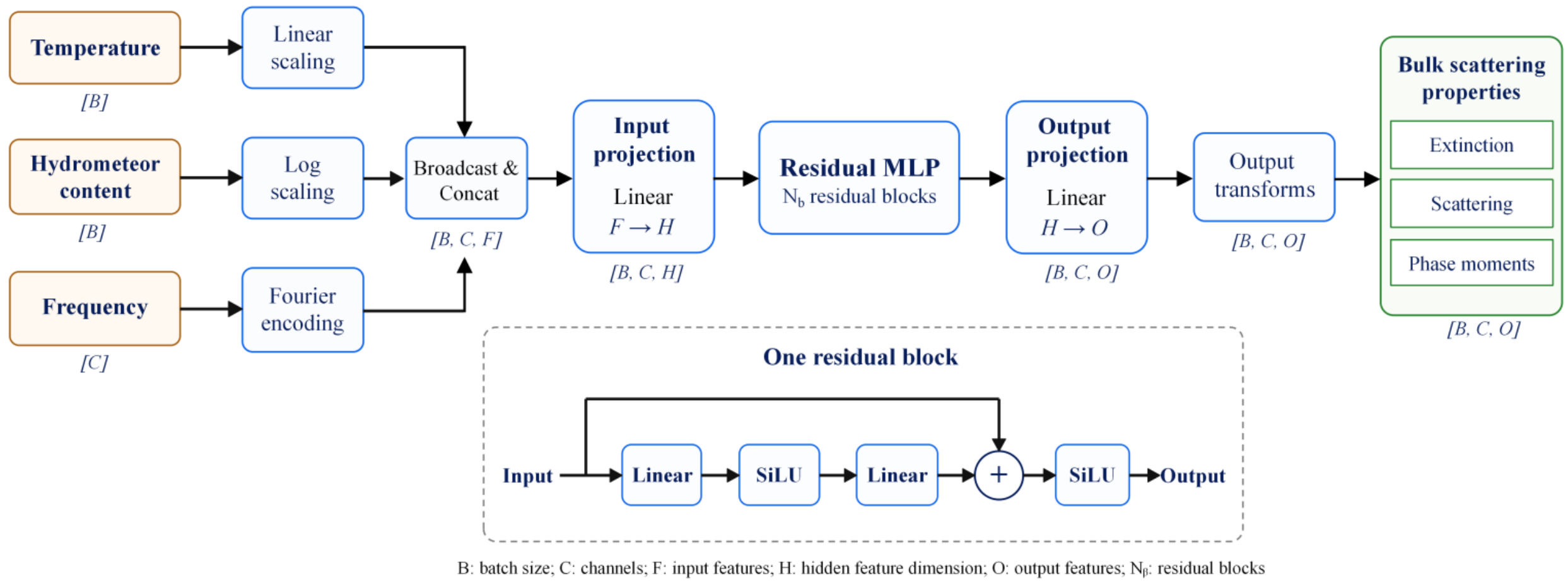

**Fig. 3.** Schematic architecture of the residual multilayer perceptron used to parameterize hydrometeor bulk scattering properties. The diagram highlights the input transformations, residual network structure, and output constraints, with the lower panel illustrating a single residual block.

## 2.4 Surface boundary

UDL-RT uses different surface emissivity parameterizations for open ocean and for land or sea ice. These parameterizations provide the surface emissivity and reflection properties required by the radiative transfer solver to represent surface thermal emission and reflected atmospheric downwelling radiation.

Over open ocean, surface emissivity is calculated using SURFEM-Ocean (Kilic et al., 2023), an NN parameterization based on the PARMIO ocean emissivity model. SURFEM-Ocean retains the dielectric and Fresnel calculations for a flat ocean surface, while NNs represent the wind-induced isotropic and azimuth-dependent contributions to microwave emissivity. The model accounts for the dependence of ocean emissivity on frequency, incidence angle, sea-surface temperature, salinity, 10 m wind speed, relative wind direction, and polarization. Over land and sea ice, UDL-RT uses the TELSEM2 microwave emissivity atlas. TELSEM2 provides monthly climatological microwave emissivities derived from satellite observations and accounts for variations with geographic location, frequency, incidence angle, and polarization (Aires et al., 2011; Wang et al., 2017).

Both surface parameterizations are accessed through a common interface that supplies the emissivity and reflection properties required by the radiative transfer solver. Users may also provide these quantities directly, allowing alternative surface parameterizations to be used without modifying the atmospheric optical-property calculations or the radiative transfer solver.

### 2.5 Radiative transfer solver

Under all-sky conditions, hydrometeor scattering redistributes microwave radiation among propagation directions and introduces multiple scattering throughout the atmospheric column. Solving the one-dimensional scattering radiative transfer equation therefore requires both an appropriate representation of the angular radiation field and an efficient treatment of radiative coupling across atmospheric layers.

For the angular treatment, low-order methods such as the $\delta$-Eddington approximation use a two-stream representation with $\delta$-scaling to account for strongly forward scattering (Joseph et al., 1976), providing high computational efficiency for fast microwave radiative transfer (Bauer et al., 2006). However, the accuracy of such low-order angular approximations can degrade under strongly scattering conditions, particularly toward higher microwave and submillimeter frequencies (Barlakas et al., 2022). Multi-stream methods instead represent the angular radiation field using a finite set of discrete propagation directions (Stamnes et al., 1988). UDL-RT adopts this multi-stream formulation, using Gauss-Legendre quadrature for angular integration and the Advanced Matrix Operator Method (AMOM; Liu and Weng, 2013) to solve the resulting coupled radiative transfer equations within each homogeneous layer. Let $N_\mu$ denote the number of quadrature directions in each hemisphere. The total number of streams, including both upward and downward directions, is

$$N_s = 2N_\mu \tag{11}$$

UDL-RT supports $N_s$ = *2*,*4*,*8*, and *16*, with four streams used as the default configuration in this study. The bulk phase-function moments derived in Section 2.3 are used to represent the scattering phase function:

$$P(\mu, \mu') = \sum_{\ell=0}^{L} (2\ell + 1)\, \beta_\ell P_\ell(\mu) P_\ell(\mu') \tag{12}$$

where $P_\ell$ is the Legendre polynomial of order $\ell$, and $L$ is the maximum retained order. In the present implementation, $L = N_s - 1$ is used by default. The hydrometeor optical-property parameterization described in Section 2.3 provides phase-function moments up to order 15, allowing these moments to be used directly for all supported stream configurations.

With the angular quadrature and phase-function representation specified, the radiative transfer equation within each homogeneous atmospheric layer reduces to a coupled system of linear ordinary differential equations. AMOM solves this single-layer problem using eigendecomposition, from which the transmission and reflection operators and thermal source vectors are derived (Liu and Weng, 2013). Let $\mathbf{I}^+$ and $\mathbf{I}^-$ denote the upward and downward radiance vectors, with $j - \frac{1}{2}$ and $j + \frac{1}{2}$ denoting the upper and lower boundaries of layer $j$. The single-layer radiative response can then be written as:

$$\boldsymbol{I}^+_{j-\frac{1}{2}} = \mathbf{T}_j \boldsymbol{I}^+_{j+\frac{1}{2}} + \mathbf{R}_j \boldsymbol{I}^-_{j-\frac{1}{2}} + \boldsymbol{S}^+_j \tag{13}$$

$$\boldsymbol{I}^-_{j+\frac{1}{2}} = \mathbf{R}_j \boldsymbol{I}^+_{j+\frac{1}{2}} + \mathbf{T}_j \boldsymbol{I}^-_{j-\frac{1}{2}} + \boldsymbol{S}^-_j \tag{14}$$

where $\mathbf{T}_j$ and $\mathbf{R}_j$ are the layer transmission and reflection operators, and $\boldsymbol{S}^+_j$ and $\boldsymbol{S}^-_j$ are the upward and downward thermal source vectors. The transmission and reflection operators are determined by the layer optical depth, single-scattering albedo, and phase function, while the thermal source vectors also depend on the layer temperature. Together, these quantities describe the radiative response of an individual atmospheric layer. Once the single-layer transmission, reflection, and source terms are obtained, the full atmospheric column is constructed recursively using the adding method (Liu and Weng, 2006). Let $\mathcal{R}_{j+1}$ denote the reflection operator of the composite system below layer $j$, and let $\boldsymbol{U}_{j+1}$ denote its upward source vector. We then define

$$\mathbf{A}_j = \left(\mathbf{E} - \mathcal{R}_{j+1}\mathbf{R}_j\right)^{-1} \tag{15}$$

where $\mathbf{E}$ is the identity matrix. Adding layer $j$ gives

$$\boldsymbol{U}_j = \mathbf{S}_j^{+} + \mathbf{T}_j\mathbf{A}_j\left(\boldsymbol{U}_{j+1} + \mathcal{R}_{j+1}\mathbf{S}_j^{-}\right) \tag{16}$$

and

$$\mathcal{R}_j = \mathbf{R}_j + \mathbf{T}_j\mathbf{A}_j\mathcal{R}_{j+1}\mathbf{T}_j \tag{17}$$

The inverse operator $\mathbf{A}_j$ accounts for repeated radiative interactions between layer $j$ and the composite atmosphere below it. The adding procedure is initialized with the surface emission and reflection boundary described in Section 2.4 and proceeds upward through the atmospheric column. At the upper boundary, the incoming downward radiance is prescribed by the isotropic cosmic microwave background. For spaceborne simulations, UDL-RT uses the upward radiance at the top of the atmosphere, while for ground-based simulations, it uses the downward radiance at the lower boundary. The resulting radiance along the specified viewing direction is then converted to Planck-equivalent TB using the inverse Planck function.

From a computational perspective, the multiple-scattering solver requires repeated eigendecompositions, linear solves, and matrix multiplications of small matrices in each atmospheric layer. Because each operation involves only a small amount of computation, repeatedly calling separate PyTorch linear algebra routines can introduce significant overhead, particularly on GPUs. We therefore implement the multiple scattering solver in C++ using Kokkos, a performance-portable programming model for parallel computing on CPUs and GPUs (Trott et al., 2022). With Kokkos, these matrix calculations are performed within a compiled solver, reducing the overhead associated with repeatedly invoking individual PyTorch routines. The solver remains differentiable and is directly integrated into PyTorch autograd.

### 2.6 Automatic differentiation

The modules described above define a mapping from atmospheric and surface variables to simulated TB:

$$\mathbf{T}_B = \mathcal{F}(\mathbf{x}; \boldsymbol{\theta}) \tag{18}$$

where $\mathbf{x}$ denotes the continuous atmospheric and surface inputs, and $\boldsymbol{\theta}$ denotes the trainable parameters of the differentiable model components. During the forward calculation, these components form a computational graph linking $\mathbf{x}$ and $\boldsymbol{\theta}$ to the simulated TBs. Reverse-mode automatic differentiation then applies the chain rule through this graph to obtain the required derivatives (Baydin et al., 2018).

The sensitivity of the simulated TBs to the model inputs is described by the Jacobian:

$$\mathbf{K} = \frac{\partial \mathbf{T}_B}{\partial \mathbf{x}} \tag{19}$$

UDL-RT can obtain sensitivities for selected channels or input variables and can construct the full Jacobian when required. In applications such as retrieval and variational data assimilation, however, optimization often requires the gradient of a scalar objective function rather than the full Jacobian itself. Given a differentiable scalar objective function $\mathcal{L}(\mathbf{T}_B, \mathbf{y})$ defined using the simulated TBs and observations $\mathbf{y}$, reverse-mode differentiation gives

$$\nabla_{\mathbf{x}}\mathcal{L} = \left(\frac{\partial \mathbf{T}_B}{\partial \mathbf{x}}\right)^{\mathrm{T}} \nabla_{\mathbf{T}_B}\mathcal{L} = \mathbf{K}^{\mathrm{T}} \nabla_{\mathbf{T}_B}\mathcal{L} \tag{20}$$

This allows the gradient with respect to the atmospheric and surface variables to be evaluated directly without explicitly constructing and storing the full Jacobian. The same computational graph can also propagate gradients to parameters of the differentiable model components:

$$\nabla_{\boldsymbol{\theta}}\mathcal{L} = \left(\frac{\partial \mathbf{T}_B}{\partial \boldsymbol{\theta}}\right)^{\mathrm{T}} \nabla_{\mathbf{T}_B}\mathcal{L} \tag{21}$$

This allows observations to be used to optimize selected trainable parameters in the gaseous absorption, hydrometeor BSP, and surface parameterizations by propagating gradients through the complete radiative transfer calculation. In this way, uncertain parameterizations can be optimized within the same differentiable framework used for forward simulation and sensitivity calculations, while the radiative transfer solver remains an explicit component of the forward model.

## 3 Evaluation of UDL-RT

UDL-RT is evaluated progressively from its individual parameterizations to the differentiable radiative transfer system. At the component level, we examine whether the gaseous absorption and hydrometeor BSP parameterizations reproduce their respective reference calculations. We then evaluate the all-sky simulations through the numerical convergence of the multi-stream radiative transfer calculations and comparisons of simulated TBs with MWHS-2 observations and RTTOV. The derivatives obtained through automatic differentiation are further verified against finite-difference calculations and compared with those from RTTOV, while computational tests quantify the efficiency of forward and gradient calculations on CPUs and GPUs.

### 3.1 Evaluation data and metrics

The evaluation dataset consists of 100,000 FY-3E MWHS-2 observation samples randomly selected from 2024 and collocated with the corresponding ERA5 atmospheric profiles. The MWHS-2 observations provide the sampling locations and viewing geometries, while ERA5 provides the atmospheric and surface states used in the radiative transfer calculations.

ERA5 is the fifth-generation global atmospheric reanalysis produced by the ECMWF. It combines an NWP model with a wide range of observations through data assimilation to provide hourly atmospheric fields at a horizontal resolution of approximately 31 km and on 137 model levels (Hersbach et al., 2020). The ERA5 model-level temperature, specific humidity, cloud fraction, and hydrometeor mixing ratios (cloud liquid water, cloud ice, rain, and snow) are used as inputs to UDL-RT. Although UDL-RT also supports graupel, it is treated diagnostically in the convection scheme of the ECMWF Integrated Forecasting System (IFS), and its mixing ratio is not archived in ERA5 (Geer et al., 2018). All required atmospheric variables are interpolated to the 54 vertical levels used by UDL-RT.

MWHS-2 onboard FY-3E is a cross-track scanning microwave sounder with 15 channels (Zhang et al., 2024). Ch2-Ch9 sample the 118.75 GHz oxygen absorption band and are primarily sensitive to atmospheric temperature, whereas Ch11-Ch15 sample the 183.31 GHz water-vapour absorption line and are primarily sensitive to atmospheric humidity. Ch1 and Ch10 are centered at 89 and 166 GHz, respectively, and are sensitive to the surface, clouds, and precipitation. The channel characteristics used in this study are summarized in Table 2.

**Table 2.** Instrument specifications of FY-3E MWHS-2

| Channel Number | Frequency (GHz) | Polarization | Bandwidth (MHz) | NEΔT (K) | Resolution (km) |
|---|---|---|---|---|---|
| 1 | 89 | QH | 1500 | 0.4 | 30 |
| 2 | 118.75 ± 0.08 | QV | 20 | 2.2 | |
| 3 | 118.75 ± 0.2 | QV | 100 | 1.0 | |
| 4 | 118.75 ± 0.3 | QV | 165 | 0.8 | |
| 5 | 118.75 ± 0.8 | QV | 200 | 0.8 | |
| 6 | 118.75 ± 1.1 | QV | 200 | 0.8 | |
| 7 | 118.75 ± 2.5 | QV | 200 | 0.8 | |
| 8 | 118.75 ± 3.0 | QV | 1000 | 0.5 | |
| 9 | 118.75 ± 5.0 | QV | 2000 | 0.5 | |
| 10 | 166.0 | QH | 1500 | 0.4 | 15 |
| 11 | 183.31 ± 1.0 | QV | 500 | 0.6 | |
| 12 | 183.31 ± 1.8 | QV | 700 | 0.6 | |
| 13 | 183.31 ± 3.0 | QV | 1000 | 0.5 | |
| 14 | 183.31 ± 4.5 | QV | 2000 | 0.5 | |
| 15 | 183.31 ± 7.0 | QV | 2000 | 0.5 | |

The agreement between UDL-RT and the reference calculations is assessed using the mean bias and standard deviation (STD) of the TB differences. For sample $i$, the brightness-temperature difference is defined as:

$$\Delta T_{\mathrm{B},i} = T_{\mathrm{B},i}^{\mathrm{sim}} - T_{\mathrm{B},i}^{\mathrm{ref}} \tag{22}$$

The mean bias and STD are calculated as:

$$\mathrm{Bias} = \frac{1}{N}\sum_{i=1}^{N} \Delta T_{\mathrm{B},i} \tag{23}$$

$$\mathrm{STD} = \sqrt{\frac{1}{N}\sum_{i=1}^{N} \left(\Delta T_{\mathrm{B},i} - \mathrm{bias}\right)^2} \tag{24}$$

Here, $T_{\mathrm{B},i}^{\mathrm{sim}}$ and $T_{\mathrm{B},i}^{\mathrm{ref}}$ denote the brightness temperatures from UDL-RT and the corresponding reference calculation, respectively, and $N$ is the number of valid samples.

### 3.2 Accuracy of the gas-absorption parameterization

Fig. 4 compares the errors in simulated brightness temperature from GasAbs-NN and the predictor-based linear-regression parameterization against the LBL reference. Realistic surface boundary conditions are retained using SURFEM-Ocean over ocean and TELSEM2 over land and sea ice, rather than prescribing a constant surface emissivity. The evaluation uses the independent set of 100,000 ERA5 profiles described in Section 3.1 under clear-sky conditions. For each comparison with the LBL reference, all other model settings are kept identical, so the resulting TB differences isolate the error introduced by the gaseous absorption parameterization.

The linear-regression parameterization yields a maximum absolute mean bias of approximately 0.011 K across the MWHS-2 channels. The STD ranges from approximately 0.006 to 0.040 K, with larger errors in the water-vapour and window channels. The magnitude and channel dependence of these errors are broadly consistent with previous evaluations of microwave transmittance regression schemes (Kan et al., 2024). In surface-sensitive channels such as Ch1, Ch9, and Ch10, relatively high atmospheric transmittance allows surface-reflected downwelling radiation to reach the sensor. Low ocean emissivity increases this reflected contribution and can amplify the impact of gaseous absorption errors (Buehler et al., 2006; Moradi et al., 2020). With a blackbody surface, the regression STDs in these channels fall below 0.01 K. The larger errors in the humidity channels may reflect the greater variability of water-vapour profiles, which makes the corresponding absorption more difficult to represent using regression predictors (Silveira et al., 2024).

Compared with linear regression, GasAbs-NN reduces the STDs across all 15 channels to approximately 0.002-0.014 K, with a maximum absolute mean bias of approximately 0.009 K. The reduction is particularly pronounced in the humidity channels, where the STDs decrease by approximately 60-85%. This improvement may partly reflect the network's ability to learn nonlinear relationships between atmospheric variables and optical depth directly from data, rather than restricting these relationships to prescribed predictor functions. The convolutional structure also preserves the vertical structure of the profiles, allowing the network to learn how variations between neighbouring layers relate to absorption. Progressively dilated convolutions further extend this vertical context by incorporating information from more distant layers into the prediction of layer optical depth.

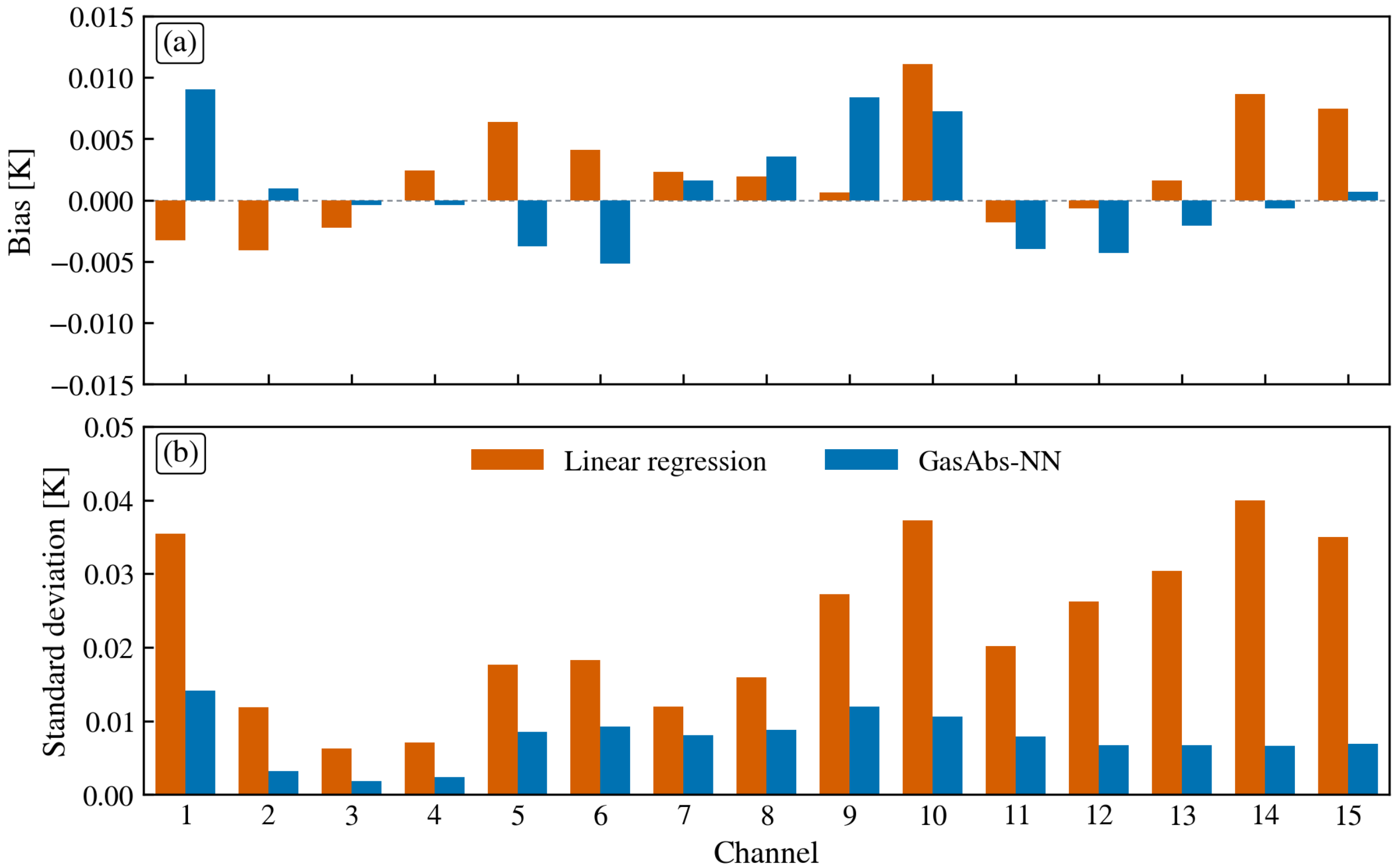

**Fig. 4.** Simulation errors of the GasAbs-NN and linear-regression gas-absorption parameterizations relative to the line-by-line (LBL) reference across the FY-3E MWHS-2 channels. Statistics are calculated from 100,000 ERA5 atmospheric profiles at the corresponding MWHS-2 viewing angles under clear-sky, gas-only conditions. Errors are defined as the difference in simulated brightness temperature between each parameterization and LBL. Panels show (a) mean bias and (b) standard deviation. Orange and blue bars denote the linear-regression and GasAbs-NN parameterizations, respectively.

### 3.3 Accuracy of the hydrometeor BSP parameterization

Direct PSD integration requires numerical integration of single-particle scattering properties over particle size and must be repeated for different atmospheric conditions and frequencies, resulting in substantial computational cost. BSP-NN replaces this procedure with NN inference and directly provides the BSPs required by the radiative transfer solver over 1.0-886.4 GHz. Compared with direct PSD integration, BSP-NN provides an approximately 79-fold speedup on the CPU, substantially reducing the computational cost of hydrometeor optical property calculations. We therefore evaluate whether the approximation introduced by BSP-NN has a significant effect on the resulting simulated TBs.

The evaluation uses the same 100,000 atmospheric profiles described in Section 3.1 and frequencies sampled at 1 GHz intervals from 1 to 886 GHz. For each profile and frequency, monochromatic gaseous optical depths are precomputed using MonoRTM. This allows the errors introduced by BSP-NN to be assessed continuously across the frequency range, rather than only at the discrete channels of a specific sensor. Fig. 5 shows that the mean differences remain close to zero across the frequency range, with absolute mean biases below approximately 0.013 K. The standard deviation varies with frequency but remains below approximately 0.03 K. The larger values at several frequencies indicate that the influence of BSP approximation on simulated TB depends on the sensitivity of the radiative transfer calculation to hydrometeor scattering at a given frequency. Overall, BSP-NN introduces only small differences in simulated TB while substantially reducing the computational cost of BSP calculations.

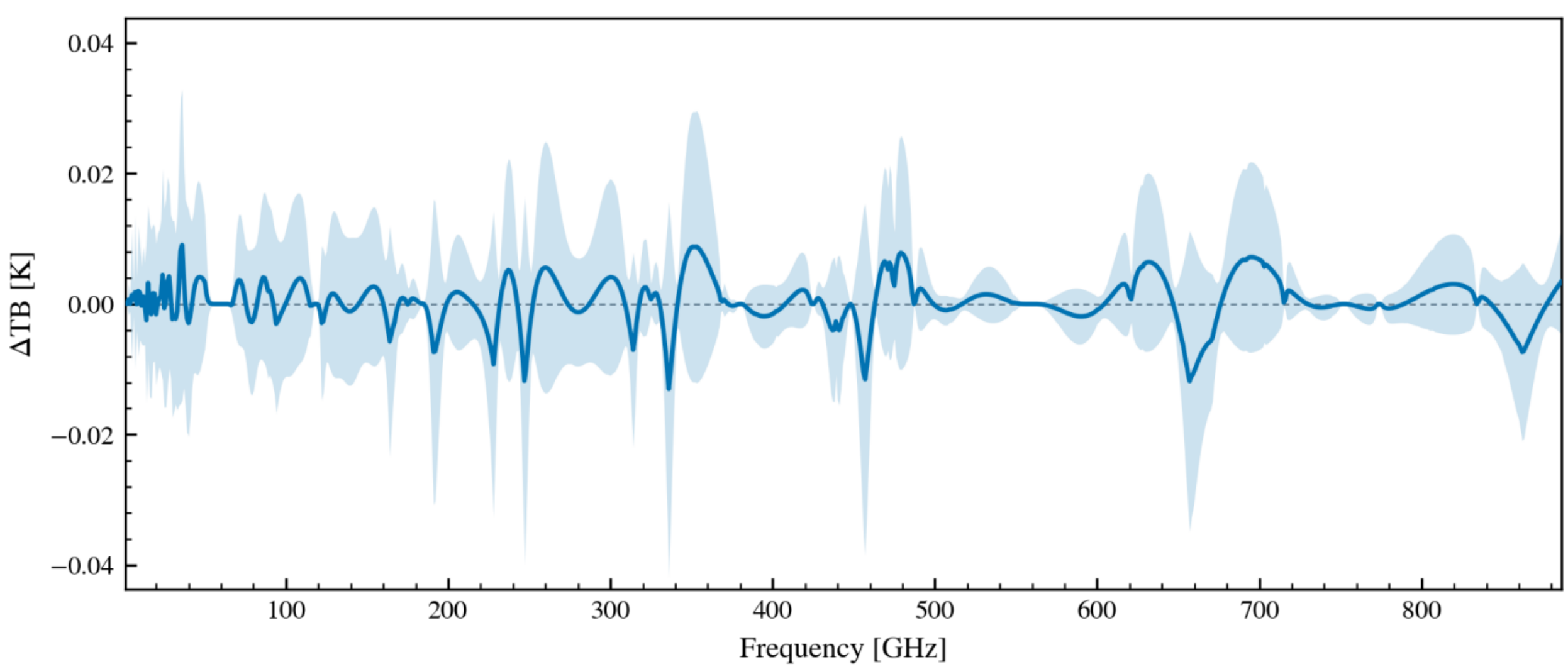

**Fig. 5.** Errors in simulated brightness temperature introduced by BSP-NN relative to direct PSD integration as a function of frequency. Statistics are calculated from 100,000 atmospheric profiles over 1.0-886 GHz with a frequency spacing of 1 GHz. Monochromatic gaseous optical depths at each frequency are precomputed using MonoRTM. Errors are defined as BSP-NN minus direct PSD integration, with all other radiative transfer settings kept identical. The solid blue line shows the mean bias, and the blue shading indicates ±1 standard deviation.

### 3.4 Evaluation of all-sky radiance simulations

The number of streams controls the angular resolution of the radiative transfer calculation and therefore affects the representation of multiple scattering. We first examine the numerical convergence of the multi-stream radiative transfer solver for four representative MWHS-2 channels, Ch1, Ch9, Ch10, and Ch15, using the test dataset described in Section 3.1. The 16-stream calculation is used as the reference. As the number of streams increases from 2 to 4 and then 8, the differences in simulated TB relative to the 16-stream reference decrease consistently across all four channels (Fig. 6). The standard deviation of the differences ranges from 0.388 to 0.972 K for two streams, from 0.045 to 0.185 K for four streams, and from 0.002 to 0.018 K for eight streams. The substantial reduction from two to four streams, followed by further improvement with eight streams, demonstrates the numerical convergence of the solver. Given the relatively small differences obtained with four streams, this configuration is adopted for the subsequent all-sky simulations.

The four-stream UDL-RT simulations are evaluated against MWHS-2 observations and compared with RTTOV v14 using its official default configuration (Hocking et al., 2025). The evaluation is restricted to ocean samples from the test dataset to reduce uncertainties associated with land and sea-ice emissivity. Fig. 7 compares the mean biases and STDs of the observation-minus-background (O-B) departures across all 15 MWHS-2 channels. The two models exhibit broadly similar channel-dependent variations in O-B STD. Ch1 and Ch10 have the largest STDs, consistent with the strong sensitivity of these window channels to the surface and hydrometeors. In contrast, temperature-sounding channels closer to the 118.75 GHz oxygen absorption line generally have smaller STDs. For the humidity-sounding channels, STDs increase from the center toward the wings of the 183.31 GHz water-vapour absorption line, consistent with their greater sensitivity to lower-tropospheric conditions and the influence of clouds and precipitation. While the two models show similar STD patterns, their mean biases differ for several channels, possibly reflecting differences in the treatment of gaseous absorption, hydrometeor optical properties, surface emissivity, and multiple scattering. For example, RTTOV uses a microwave gaseous absorption parameterization derived from AMSUTRAN calculations (Turner et al., 2019), whereas GasAbs-NN is trained against channel optical depths generated by MonoRTM. Overall, UDL-RT yields O-B biases and STDs of similar magnitude to those from RTTOV across the evaluated MWHS-2 channels, indicating comparable performance in all-sky forward simulations.

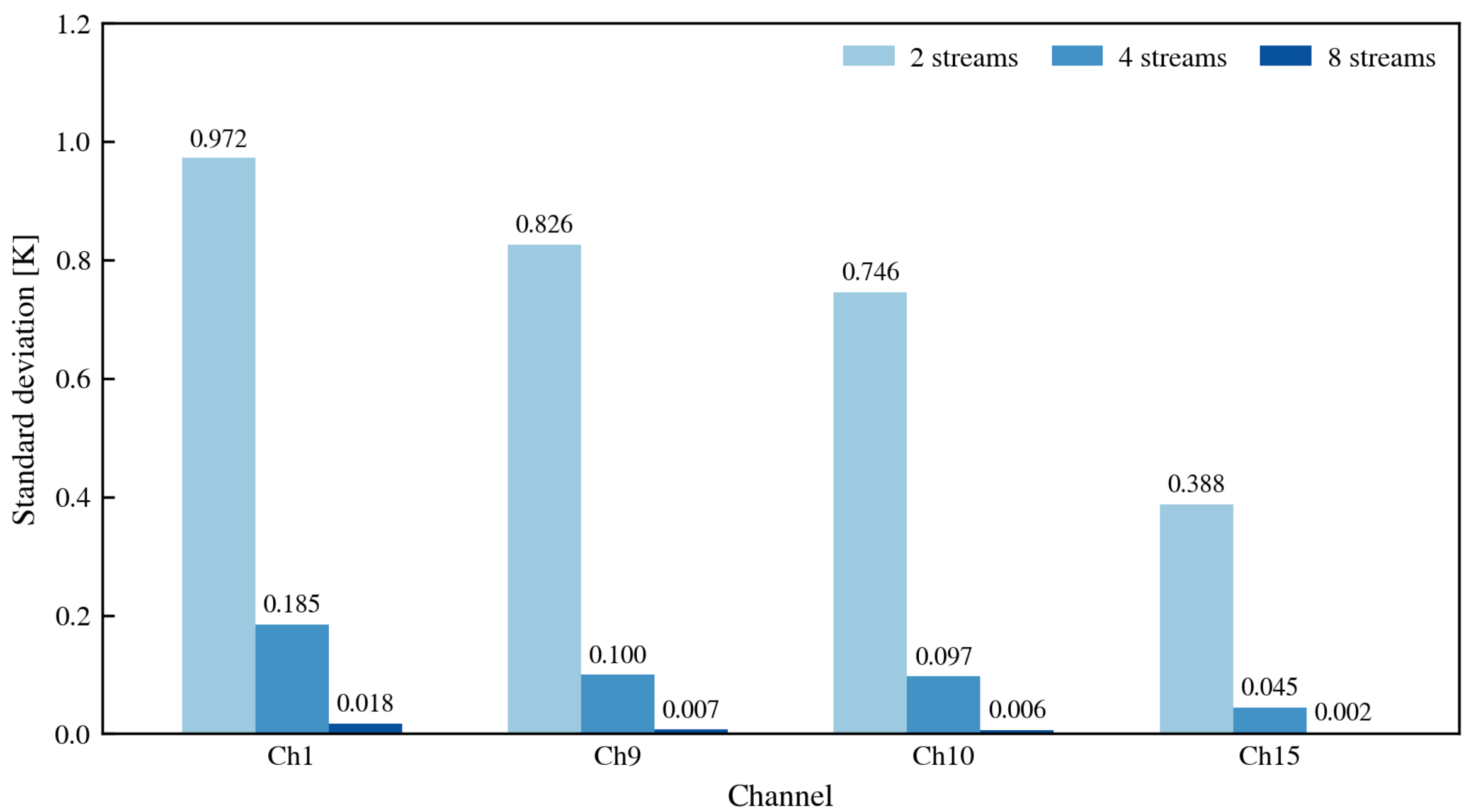

**Fig. 6.** Numerical convergence of the all-sky brightness-temperature calculations with increasing numbers of streams. The standard deviation of the brightness-temperature differences relative to the 16-stream solution is calculated from 100,000 atmospheric profiles for selected FY-3E MWHS-2 channels. Results are shown for 2-, 4-, and 8-stream configurations.

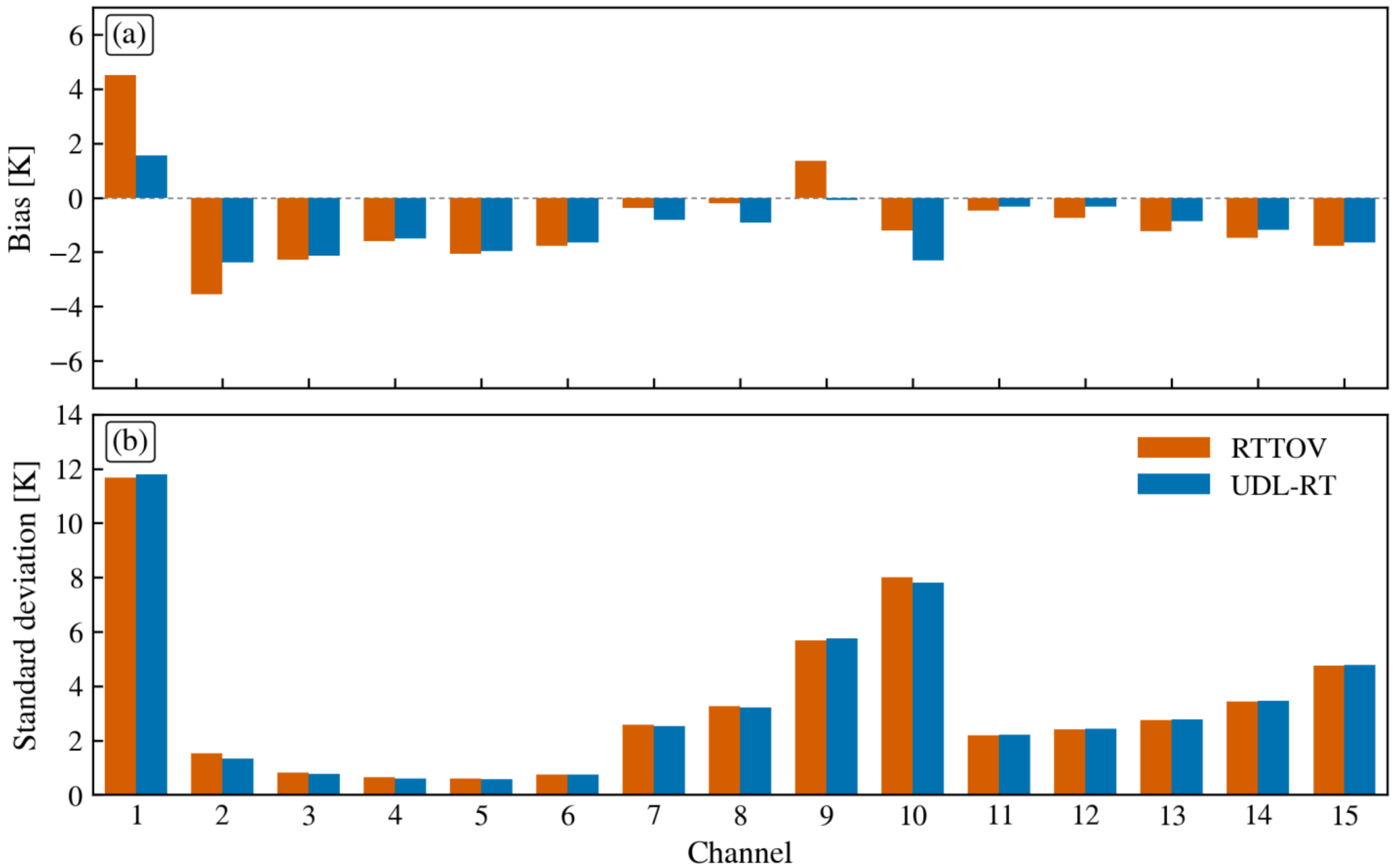

**Fig. 7.** (a) Mean bias and (b) standard deviation of all-sky O-B departures across the 15 FY-3E MWHS-2 channels. Orange and blue bars denote RTTOV and four-stream UDL-RT, respectively. Statistics are calculated from 66,299 open-ocean profiles valid for both models.

### 3.5 Verification of all-sky Jacobians

Jacobians describe how the simulated TB in each channel responds to a small perturbation in an atmospheric variable at a given level. Their signs indicate whether the perturbation increases or decreases brightness temperature, while their vertical profiles identify the layers to which each channel is most sensitive under the specified atmospheric conditions. Evaluating these derivatives therefore involves two complementary questions: whether they are computed correctly and whether they reproduce the expected channel-dependent atmospheric responses.

The Jacobians obtained by automatic differentiation are first verified against central finite differences. Finite differences are commonly used for gradient verification in differentiable physical models (Davenport et al., 2026; Kallel et al., 2025; Logan et al., 2020). A total of 1,000 atmospheric profiles are randomly selected from the evaluation dataset described in Section 3.1, covering all 15 MWHS-2 channels and 54 atmospheric layers. For each atmospheric variable, perturbations are applied layer by layer while all other inputs are held fixed. The resulting finite-difference estimates are used as the numerical reference. The relative difference between the automatic differentiation and finite-difference Jacobians is defined as:

$$\varepsilon_{n,i,j} = \frac{\left|K_{n,i,j}^{\mathrm{AD}} - K_{n,i,j}^{\mathrm{FD}}\right|}{\left|K_{n,i,j}^{\mathrm{FD}}\right|} \tag{25}$$

where $n$, $i$, and $j$ denote the profile, channel, and vertical-layer indices, respectively. $K^{\mathrm{AD}}$ and $K^{\mathrm{FD}}$ denote the Jacobians obtained from automatic differentiation and finite differences. For each variable, the 99th percentile (P99) of the relative differences is calculated over all profiles, channels, and vertical layers. The P99 values are 0.0523% and 0.0101% for the temperature and humidity Jacobians, respectively, and no larger than 0.000710% for cloud liquid water, cloud ice, rain, and snow. These results confirm the numerical accuracy of the Jacobians computed by automatic differentiation in UDL-RT.

The all-sky Jacobians from UDL-RT are then compared with those from RTTOV. Given the different parameterizations and radiative transfer solvers used by the two models, the comparison focuses on the overall vertical structure of the Jacobians rather than exact agreement in magnitude. As shown in Fig. 8, the temperature Jacobians reflect the different vertical sensitivities of the selected channels. Ch6 shows the strongest temperature sensitivity in the middle and upper troposphere, whereas Ch1, Ch10, and Ch15 are primarily sensitive to the lower troposphere. The humidity Jacobians also exhibit clear channel-dependent vertical structures. Notably, the UDL-RT humidity Jacobians vary smoothly with height and gradually approach zero in the upper atmosphere, where water vapour concentrations are small. They do not exhibit the oscillations in the upper atmosphere previously reported for NN emulators (Li et al., 2024a).

The hydrometeor Jacobians also show similar vertical structures in UDL-RT and RTTOV. For cloud liquid water, Ch1 and Ch10 exhibit positive sensitivities in the lower troposphere, as liquid clouds act as warm emitters against the ocean background with relatively low emissivity. In contrast, Ch6 exhibits negative sensitivities, as increased cloud liquid water shifts the weighting function toward higher and colder atmospheric layers. The snow Jacobians are predominantly negative across all four channels, with larger magnitudes in Ch10 and Ch15 than in Ch1 and Ch6, reflecting stronger ice-particle scattering at higher frequencies. Overall, the UDL-RT all-sky Jacobians capture the main vertical sensitivity structures of microwave TBs to temperature, humidity, and hydrometeors. Together with the finite-difference verification, these results support the numerical accuracy and physical consistency of the UDL-RT Jacobians and provide a basis for the gradient-based parameter optimization described in Section 4.

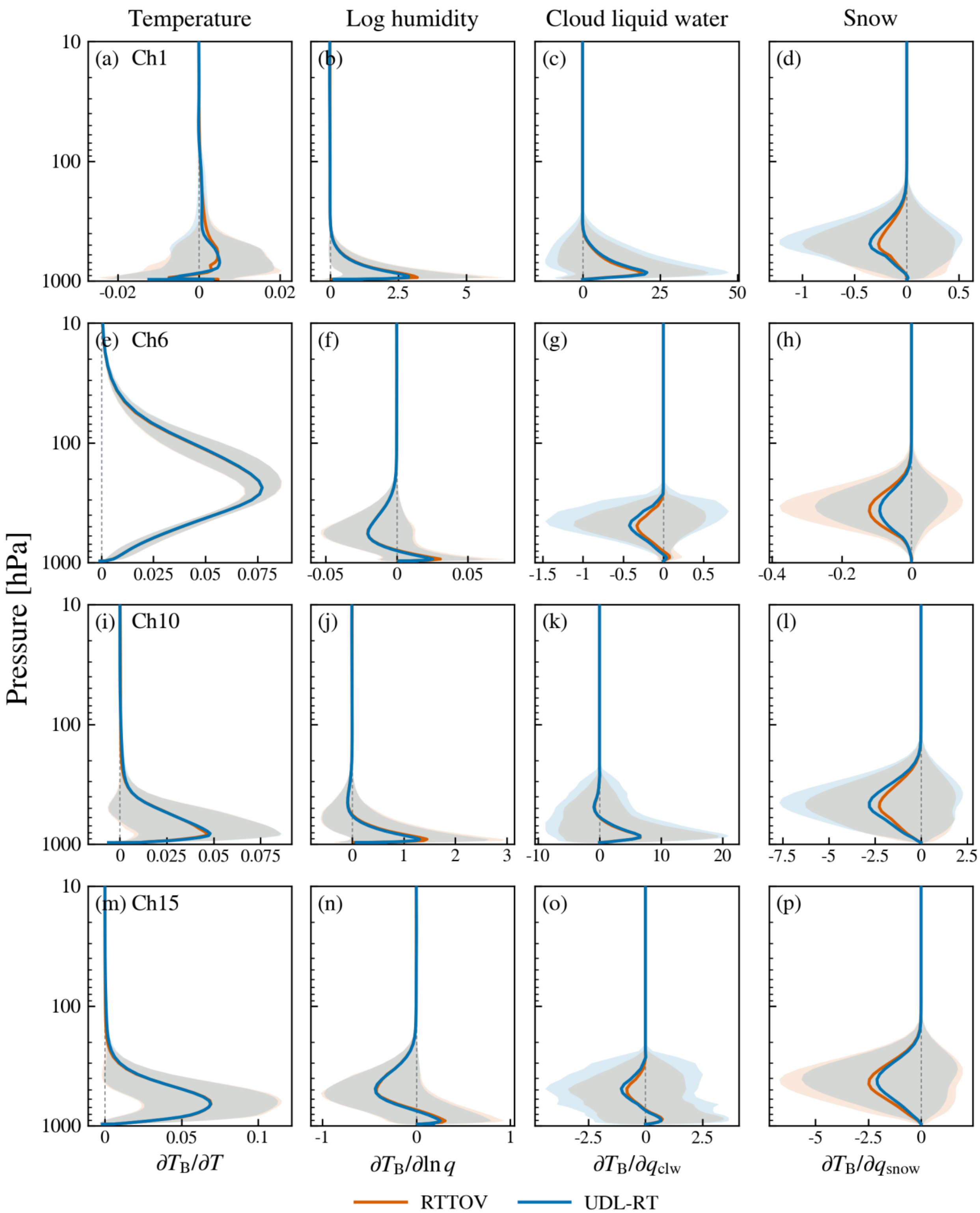

**Fig. 8.** Vertical profiles of all-sky brightness-temperature Jacobians from UDL-RT and RTTOV for FY-3E MWHS-2 Ch1, Ch6, Ch10, and Ch15. Columns show the sensitivities to temperature, logarithmic specific humidity, cloud liquid water, and snow, respectively, while rows correspond to the four channels. Solid lines show the mean Jacobians, with UDL-RT in blue and RTTOV in orange. Shaded areas indicate ±1 standard deviation around the mean.

### 3.6 Computational efficiency

Table 3 summarizes the computational performance of UDL-RT for forward and gradient calculations over 100,000 profiles and all MWHS-2 channels from the evaluation dataset described in Section 3.1. Two UDL-RT configurations, Baseline and NN-Parameterized, are considered to isolate the additional computational cost introduced by GasAbs-NN and BSP-NN. The Baseline uses the linear gas-absorption parameterization together with the RTTOV hydrotable for hydrometeor BSPs, whereas the NN-Parameterized configuration replaces these components with GasAbs-NN and BSP-NN, respectively. Both configurations use the same SURFEM-Ocean parameterization for ocean surface emissivity. The performance tests were conducted on an Intel Xeon Gold 6458Q processor (32 cores) for CPU calculations and a single AMD Instinct MI300X for GPU calculations. In UDL-RT, the NN modules used FP32 precision, while all other components used FP64. After an initial warm-up, each test was repeated ten times, and the median runtimes are reported in Table 3.

Among the UDL-RT configurations, the 2-stream Baseline is the most appropriate for comparison with RTTOV on the CPU, since the delta-Eddington scheme used in RTTOV is a variant of the two-stream radiative transfer approximation (Geer and Baordo, 2014; Meador and Weaver, 1980). The forward runtime is 2.418 s for UDL-RT and 2.865 s for RTTOV, while the corresponding gradient runtimes are 10.879 and 7.354 s. These results indicate that the batched parallel implementation of UDL-RT can effectively utilize multicore CPU resources, enabling computational performance comparable to that of the operational fast RTMs.

Replacing the conventional parameterizations with GasAbs-NN and BSP-NN increases the computational cost on the CPU. For the 4-stream case, the forward and gradient runtimes are 8.809 and 52.587 s, respectively, approximately three times those of the Baseline configuration.

GasAbs-NN applies vertical convolution operations, while BSP-NN performs MLP inference for atmospheric layers containing hydrometeors. These operations require more extensive matrix computations than the linear gas-absorption calculation and precomputed hydrotable interpolation used in Baseline.

GPU execution substantially reduces this additional cost. For the 2-stream NN-Parameterized configuration, the forward and gradient runtimes decrease from 8.607 and 49.445 s on the CPU to 0.526 and 2.691 s on the GPU, corresponding to speedups of approximately 16.4 and 18.4, respectively. This acceleration is consistent with the computational characteristics of convolutional and fully connected layers, which involve large numbers of regular multiply-accumulate operations that can be efficiently executed in parallel on GPUs (Chetlur et al., 2014; Nickolls and Dally, 2010). The GPU timings further show that the additional runtime associated with the NN parameterizations accounts for a progressively smaller fraction of the total runtime as the number of streams increases. At 16 streams, the gradient runtimes of Baseline and NN-Parameterized are 38.715 and 39.473 s, respectively, showing little difference between the two configurations.

Overall, UDL-RT achieves CPU performance comparable to RTTOV, while GPU execution substantially reduces both forward and gradient runtimes. For the 2-stream configuration, Baseline on the GPU is approximately 13.8 times faster than RTTOV for forward calculations and 3.8 times faster for gradient calculations. With the NN parameterizations enabled, the corresponding speedups remain approximately 5.4 and 2.7, respectively. These results demonstrate that the batched implementation and GPU acceleration of UDL-RT substantially improve the efficiency of forward and gradient calculations for all-sky microwave radiances.

**Table 3.** Computational performance of RTTOV and UDL-RT

| Model | Streams | CPU | | GPU | |
|---|---|---|---|---|---|
| | | Forward (s) | Backward (s) | Forward (s) | Backward (s) |
| RTTOV | 2 | 2.865 | 7.354 | — | — |
| Baseline | 2 | 2.418 | 10.879 | 0.207 | 1.956 |
| Baseline | 4 | 2.745 | 14.969 | 0.388 | 2.302 |
| Baseline | 8 | 3.558 | 26.505 | 1.200 | 7.755 |
| Baseline | 16 | 7.178 | 67.710 | 6.919 | 38.715 |
| NN-Parameterized | 2 | 8.607 | 49.445 | 0.526 | 2.691 |
| NN-Parameterized | 4 | 8.809 | 52.587 | 0.703 | 3.021 |
| NN-Parameterized | 8 | 9.766 | 64.917 | 1.523 | 8.496 |
| NN-Parameterized | 16 | 13.682 | 107.004 | 7.369 | 39.473 |

# 4 Optimization of learnable parameterizations

Hydrometeor optical properties remain a major source of uncertainty in all-sky microwave simulations, especially for frozen particles whose shapes and size distributions are poorly constrained (Geer, 2021). Here, we use the snow BSP-NN as a test case to demonstrate the feasibility of optimizing learnable parameterizations using satellite observations within UDL-RT.

## 4.1 Experimental setup

Fig. 9 illustrates how satellite observations are used to optimize hydrometeor BSP parameterizations within UDL-RT. In the forward calculation, the BSP-NNs predict hydrometeor optical properties from the atmospheric state and channel frequency. The physical radiative transfer solver then uses these properties, together with gaseous optical depths and surface boundary conditions, to simulate multi-channel TBs. The mismatch between simulated and observed TBs is quantified by a loss function. Because the forward calculation and loss function are differentiable, automatic differentiation can be used to compute the loss gradients with respect to the trainable network parameters (Eq. (21)). These gradients guide iterative parameter updates to minimize the loss.

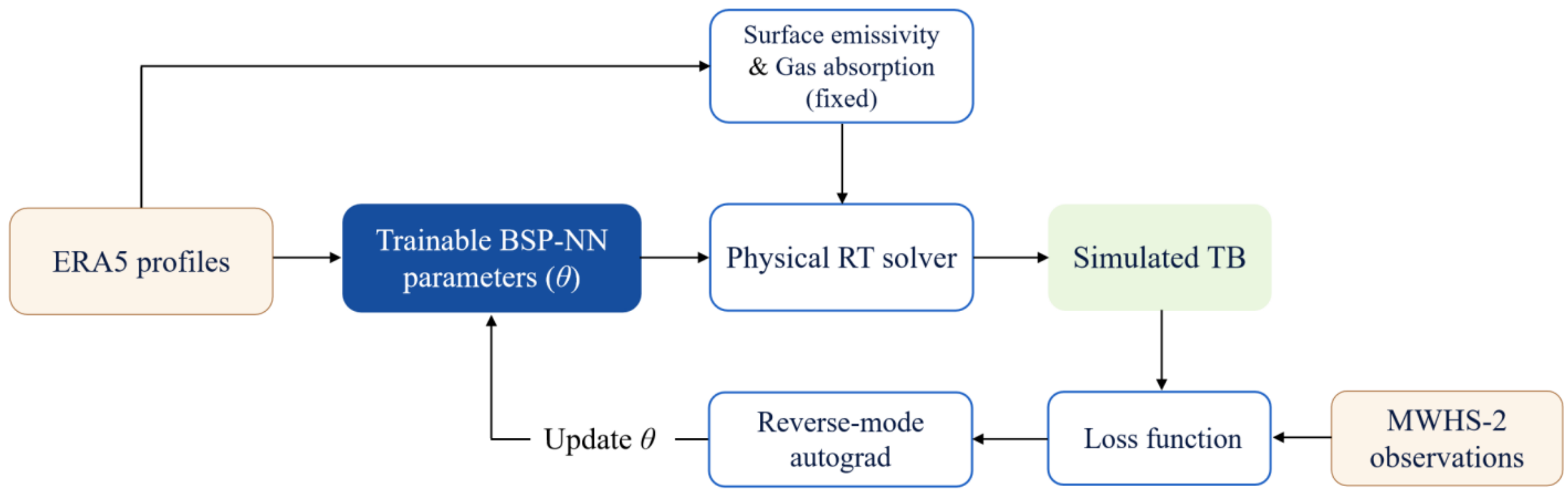

**Fig. 9.** Observation-constrained optimization of the BSP-NN within UDL-RT. ERA5 profiles provide the atmospheric and surface inputs, and MWHS-2 observations constrain the network through a loss based on TB distributions. Reverse-mode automatic differentiation propagates the loss gradients to the trainable snow parameters. The other hydrometeor parameterizations, gaseous absorption, surface emissivity, and the physical radiative transfer solver remain fixed.

The optimization uses FY-3E MWHS-2 Ch1 and Ch5-Ch15 to constrain hydrometeor BSP across different atmospheric layers and cloud conditions. Ch2-Ch4 are excluded because their weighting functions are concentrated mainly in the stratosphere, limiting their sensitivity to hydrometeors. Only ocean scenes between 50°S and 50°N are retained to reduce uncertainties associated with land and sea-ice emissivity. Superobservations are formed by averaging TBs within a 30 km radius to reduce spatial-scale differences between the observations and ERA5 (Duncan et al., 2024). Constrained bias correction is applied before optimization to reduce systematic differences between observed and simulated TBs, with coefficients estimated from clear-sky samples (Han and Bormann, 2016; Li and Han, 2024). After quality control, approximately 10 million samples from 2023 are randomly selected for training, while the validation and test sets each contain approximately 5 million randomly selected samples from 2024.

In this experiment, only the snow BSP-NN is optimized, given the substantial uncertainty in snow microphysics and the high sensitivity of simulated microwave TBs to its representation (Geer and Baordo, 2014; Gorgucci et al., 2021). The network is initialized from the pretrained model described in Section 2.3, while the weights of the BSP-NNs for cloud liquid water, cloud ice, and rain remain fixed throughout the optimization. Because ice crystal habits depend on temperature, ice supersaturation, and atmospheric pressure (Bailey and Hallett, 2009; Hueholt et al., 2022), the pretrained snow BSP-NN is extended with feature-wise linear modulation (FiLM; Perez et al., 2018) layers. These layers use pressure and specific humidity as conditioning inputs to scale and shift the hidden network features (Li et al., 2026).

Accurately predicting the location and intensity of clouds and precipitation remains challenging for NWP models, and spatial mismatches can introduce a double-penalty effect in TB comparisons (Geer and Baordo, 2014; Geer, 2021). To mitigate this effect, the loss is defined as the mean Jensen-Shannon (JS) divergence between observed and simulated TB distributions over the selected channels and 10° × 10° latitude-longitude regions (Li et al., 2026):

$$\mathcal{L} = \frac{1}{2N_r N_c}\sum_{r=1}^{N_r}\sum_{i=1}^{N_c}\sum_{k=1}^{N_{\mathrm{bin}}}\left[p^{\mathrm{o}}_{r,i,k}\ln\frac{p^{\mathrm{o}}_{r,i,k}}{\overline{p}_{r,i,k}} + p^{\mathrm{b}}_{rik}\ln\frac{p^{\mathrm{b}}_{r,i,k}}{\overline{p}_{r,i,k}}\right], \quad (26)$$

where $r$, $i$, and $k$ index regions, selected channels, and TB bins, respectively. Here, $N_r$ is the number of regions, $N_c$ is the number of selected channels, and $N_{\mathrm{bin}}$ is the number of TB bins. The normalized probabilities $p^{\mathrm{o}}$ and $p^{\mathrm{b}}$ represent observations and simulations, with $\overline{p} = \frac{p^{\mathrm{o}} + p^{\mathrm{b}}}{2}$. Both distributions are estimated using Gaussian kernel density estimation with a bandwidth of 5 K and normalized over common TB bins, making the loss differentiable with respect to the simulated TBs.

The snow BSP-NN is optimized with Adam at a learning rate of $10^{-5}$ for 50 epochs. The model with the lowest validation loss is used in the Optimized experiment, while the Control experiment uses the pretrained snow parameterization. Both experiments use the four-stream configuration.

### 4.2 Evaluation of the optimized parameterization

The optimized snow parameterization reduces the systematic warm biases and the asymmetry of the O-B distributions in the four representative channels (Fig. 10). The largest reductions in bias occur in Ch10 and Ch15, where the mean O-B departures change from -1.57 to -0.40 K and from -0.85 to -0.10 K, respectively. Their skewness values also move toward zero, from -6.25 to -2.19 in Ch10 and from -7.85 to -3.86 in Ch15. Ch1 and Ch6 also show reductions in the magnitudes of both bias and skewness. These changes are favorable for all-sky data assimilation, where errors are commonly assumed to be unbiased and Gaussian.

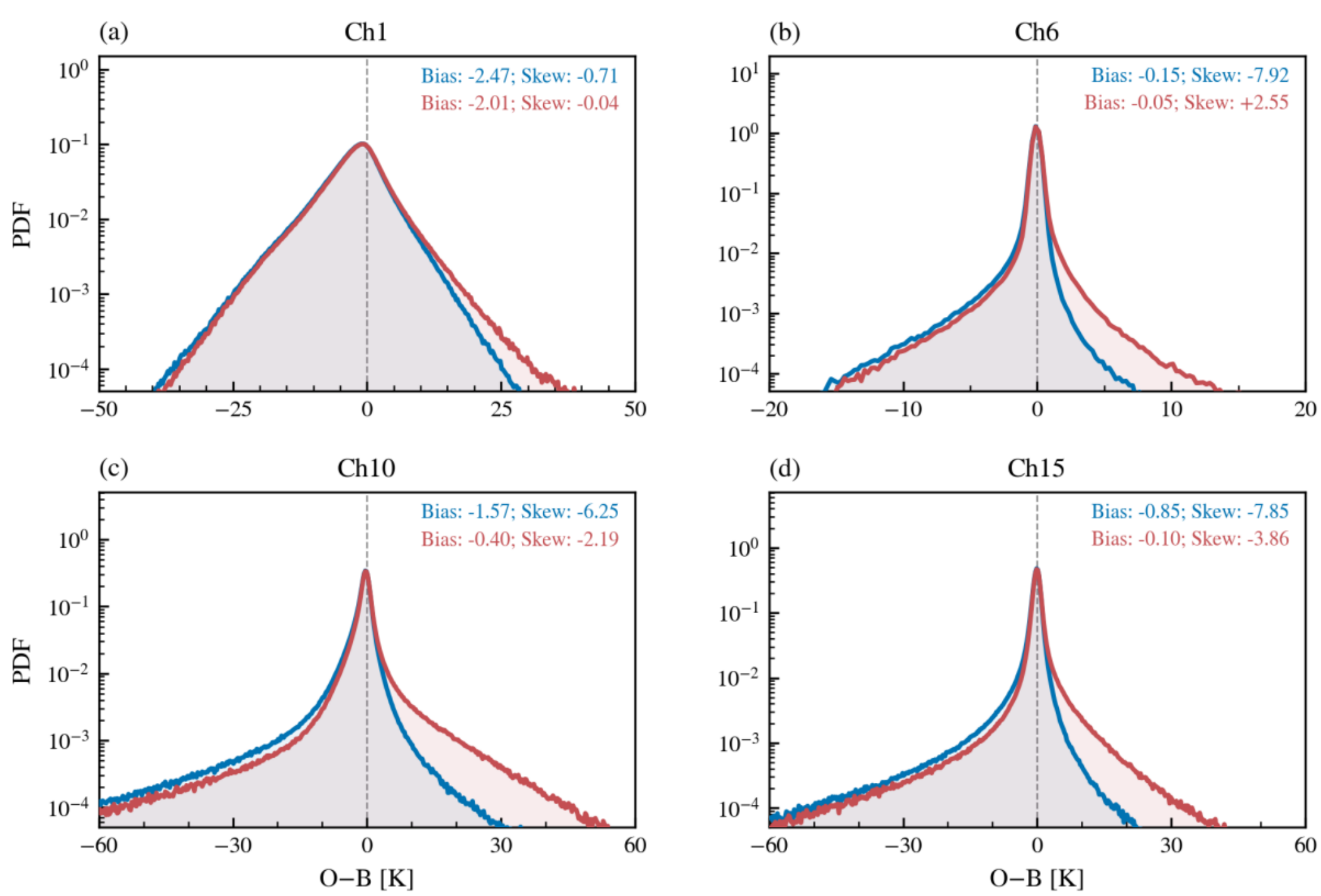

**Fig. 10.** Probability density functions of O-B departures for FY-3E MWHS-2 (a) Ch1, (b) Ch6, (c) Ch10, and (d) Ch15 in the test dataset. Blue and red curves denote the Control and Optimized experiments, respectively.

Among the selected MWHS-2 channels, Ch10 is the most sensitive to snow scattering and is therefore used to examine the regional effects of optimization. As shown in Fig. 11, the Control experiment exhibits predominantly negative O-B biases, reaching several kelvin over the western Pacific warm pool, the intertropical convergence zone, and the midlatitude oceans. By contrast, smaller biases occur over the subtropical southeastern Pacific, where subsidence suppresses deep convection (Rahn and Garreaud, 2010). After optimization, the widespread negative biases are substantially reduced, with mean departures approaching zero over most of the ocean domain. JS divergence also decreases from around $10^{-2}$ to $10^{-3}$ or lower in many regions, indicating closer agreement between the simulated and observed TB distributions.

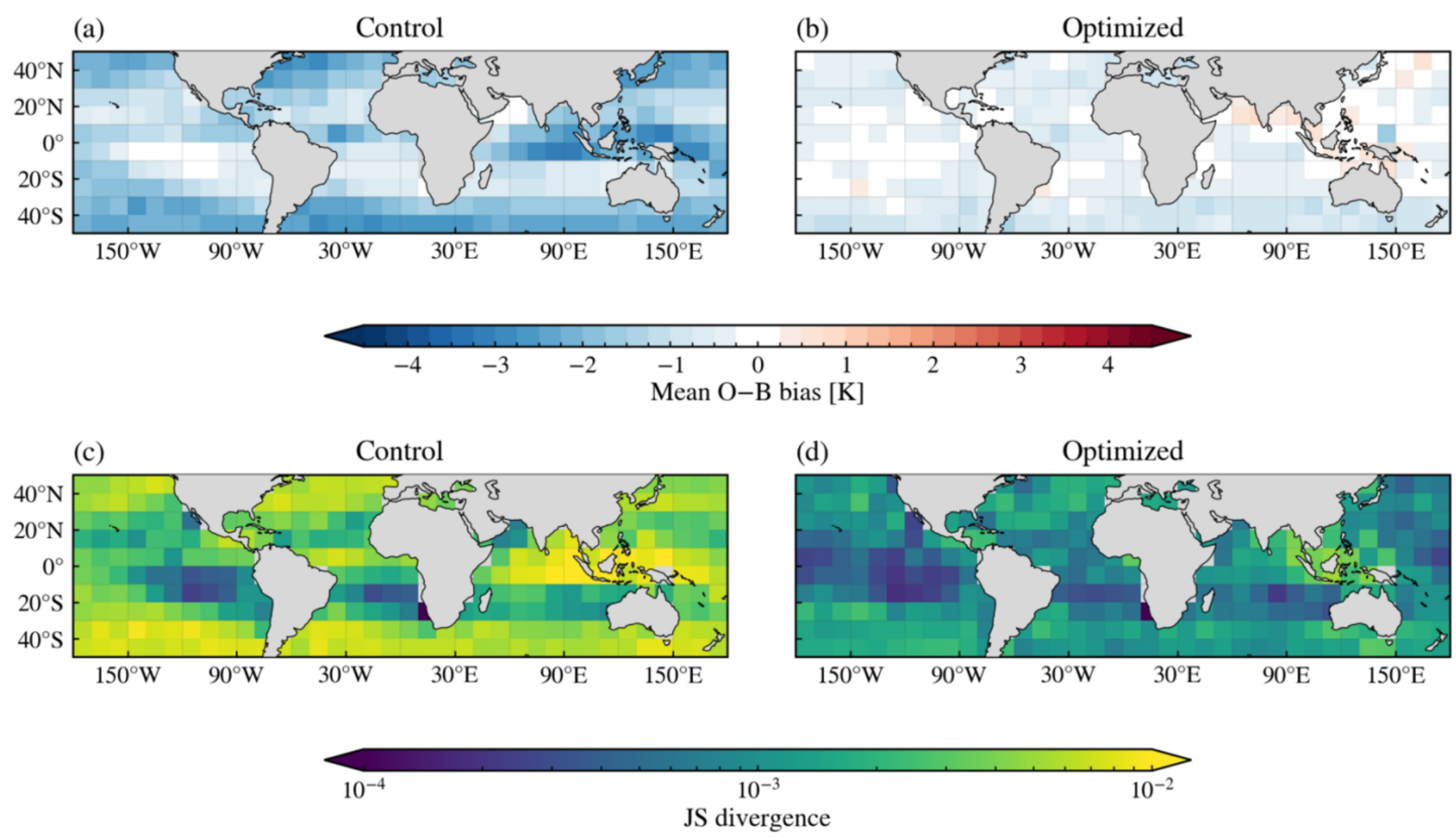

**Fig. 11.** Geographical distributions of (a, b) mean O-B bias and (c, d) JS divergence for FY-3E MWHS-2 Ch10. The left and right columns show the Control and Optimized experiments, respectively. Statistics are calculated from the test dataset in 10° × 10° latitude-longitude regions over ocean between 50°S and 50°N.

We further evaluate the optimized snow parameterization under strong scattering conditions using Ch10 observations of Typhoon Ragasa at approximately 09:00 UTC on 21 September 2025 (Fig. 12). The observations show pronounced TB depressions in the inner spiral cloud bands, broadly corresponding to the region of high snow water path in ERA5. The Control experiment captures the broad cloud-band structure but produces warmer TBs than observed in this region. In the Optimized experiment, the TB depressions become deeper and more extensive, indicating a stronger simulated scattering signal and better agreement with the observed TB depressions. Together, these results demonstrate that neural networks coupled to UDL-RT can be trained against satellite observations using automatic differentiation.

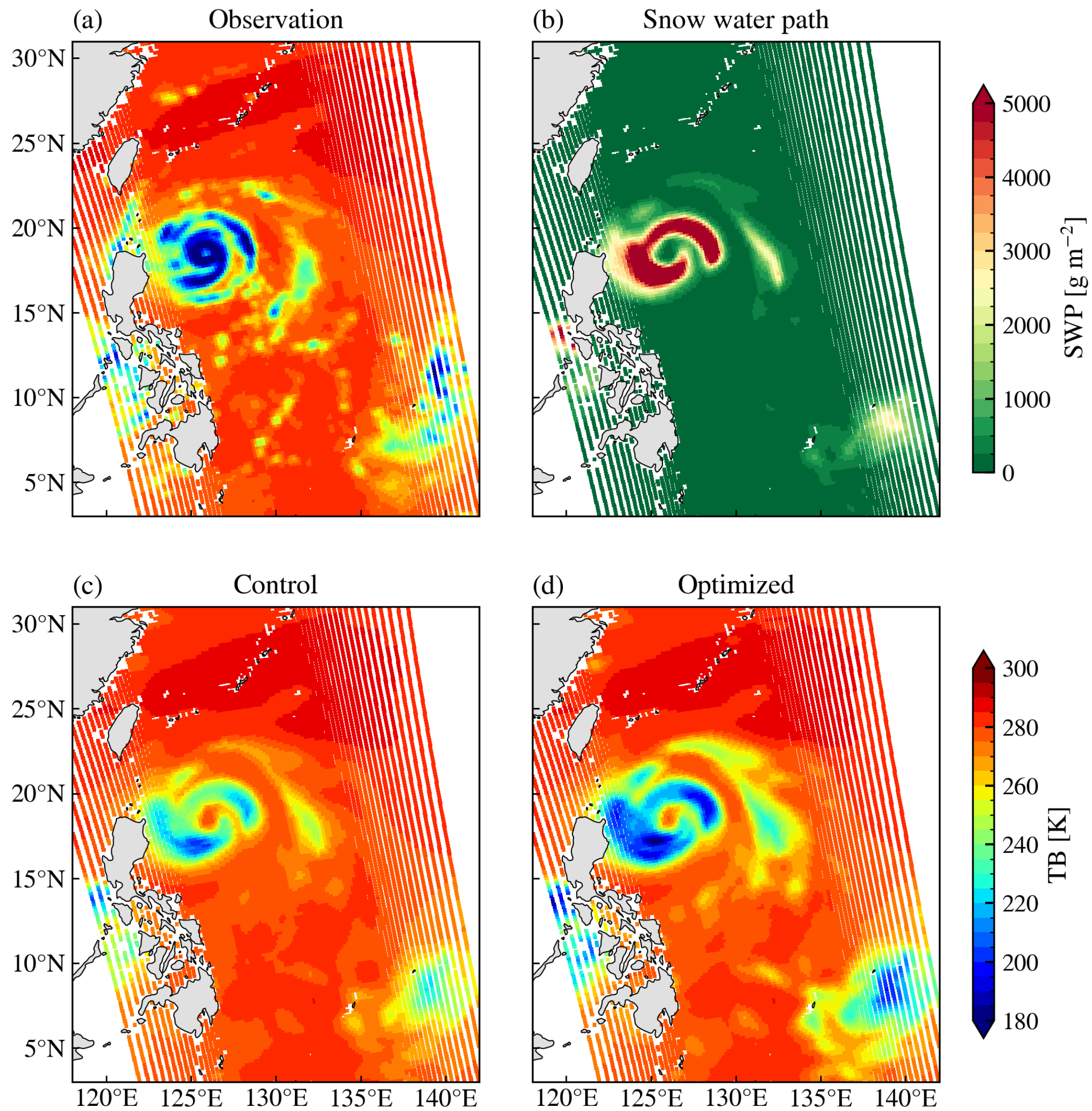

**Fig. 12.** FY-3E MWHS-2 Ch10 observations and simulations for Typhoon Ragasa during 08:49-08:57 UTC on 21 September 2025. Panels show (a) observed brightness temperature, (b) ERA5 snow water path, and brightness temperatures simulated using (c) the Control and (d) the Optimized snow BSP parameterizations.

## 5 Conclusions and discussion

We developed UDL-RT, a PyTorch-based differentiable RTM for all-sky microwave observations that integrates an explicit physical solver with learnable NN parameterizations. The model uses AMOM to solve multi-stream radiative transfer within individual atmospheric layers and the adding method to account for radiative coupling and multiple scattering between layers. The parameterizations for gaseous absorption, hydrometeor BSPs, and ocean surface emissivity are coupled to the solver within a unified differentiable computational graph. This formulation supports radiance simulation, sensitivity analysis, and observation-constrained optimization of NN parameterizations.

The NN parameterizations reproduced the reference calculations with small differences in simulated TBs. GasAbs-NN yielded STDs of approximately 0.002-0.014 K for TB differences relative to the LBL reference across the MWHS-2 channels, consistently below those obtained with the predictor-based linear-regression parameterization. Replacing direct PSD integration with BSP-NN introduced TB differences with STDs below 0.03 K over 1-886 GHz. The multi-stream solver also showed numerical convergence toward the 16-stream reference as the number of streams increased. Comparisons with MWHS-2 observations over ocean yielded O-B biases and STDs of similar magnitude to those obtained with RTTOV, supporting the model's ability to simulate all-sky microwave TBs.

In addition to its forward simulation capabilities, UDL-RT provided numerically accurate Jacobians for temperature, humidity, and hydrometeors, as verified by close agreement with centered finite differences. The vertical Jacobian profiles also showed smooth variations with height and reflected the different sensitivities of the MWHS-2 channels to atmospheric layers. Temperature sensitivities were strongest in the middle and upper troposphere for Ch6 and in the lower troposphere for window channels. The signs and vertical distributions of the humidity and hydrometeor Jacobians were consistent with the expected radiative responses of the individual channels. In UDL-RT, forward and gradient calculations share the same computational graph, allowing model updates to be reflected directly in the gradients. This eliminates the need to develop and maintain separate tangent-linear and adjoint models as the forward model evolves.

Batched computation and GPU acceleration improved the efficiency of forward and gradient calculations. UDL-RT required 2.418 s for CPU forward calculations using two streams and conventional parameterizations, compared with 2.865 s for RTTOV. On the GPU, UDL-RT achieved forward and gradient speedups of approximately 14 and 4, respectively, relative to RTTOV. Introducing GasAbs-NN and BSP-NN increased computational cost, but their regular matrix operations were well suited to GPU parallelism. With these NN parameterizations, the two-stream GPU forward and gradient runtimes were 0.526 and 2.691 s, respectively. GPU acceleration therefore makes learnable parameterizations more practical for applications requiring repeated forward and gradient calculations over large numbers of atmospheric profiles, including retrieval, data assimilation, and parameterization optimization.

UDL-RT also supports the optimization of NN parameterizations against observations through automatic differentiation. This capability was demonstrated by optimizing the pretrained snow BSP-NN using multi-channel MWHS-2 observations. Loss gradients were propagated through the physical radiative transfer solver to update the network parameters. The optimized parameterization reduced systematic warm biases and improved agreement between observed and simulated TB distributions across much of the ocean domain. The tropical cyclone case also showed closer agreement between simulated and observed TB depressions. These results demonstrate that UDL-RT enables pretrained NN parameterizations to be fine-tuned using radiance observations.

Although this study demonstrated the capabilities of UDL-RT in all-sky microwave simulation, sensitivity analysis, and parameter optimization, several aspects require further investigation. The present evaluation focused on satellite observations, while the model's capability to simulate ground-based downwelling radiances remains to be systematically assessed. Future work will use ground-based microwave radiometer measurements to evaluate downwelling TBs and examine the associated Jacobians. The general formulation of the radiative transfer solver also provides a basis for extending UDL-RT to the longwave infrared spectral region. Such an extension could retain the existing AMOM and adding framework, with development focused on infrared parameterizations of gaseous absorption, hydrometeor optical properties, and surface emission and reflection. Microwave and infrared observations could then provide complementary constraints on hydrometeor BSPs and the underlying microphysical assumptions, such as particle shape and size distribution. In terms of applications, the differentiable formulation enables efficient gradient computation for inverse problems. Future work will evaluate UDL-RT through atmospheric retrieval and cycled data assimilation experiments, assessing its impact on retrieval accuracy, analysis quality, and forecast skill.